# The Capacity Region of $p$-Transmitter/$q$-Receiver Multiple-Access Channels with Common Information

A. Haghi *Student Member, IEEE,* R. Khosravi-Farsani *Student Member, IEEE,* M. R. Aref, and F. Marvasti *Senior Member, IEEE*

**Abstract:** This paper investigates the capacity problem for some multiple-access scenarios with cooperative transmitters. First, a general Multiple-Access Channel (MAC) with common information, i.e., a scenario where $p$ transmitters send private messages and also a common message to $q$ receivers and each receiver decodes all of the messages, is considered. The capacity region of the discrete memoryless channel is characterized. Then, the general Gaussian fading MAC with common information wherein partial Channel State Information (CSI) is available at the transmitters (CSIT) and perfect CSI is available at the receivers (CSIR) is investigated. A coding theorem is proved for this model that yields an exact characterization of the throughput capacity region. Finally, a two-transmitter/one-receiver Gaussian fading MAC with conferencing encoders with partial CSIT and perfect CSIR is studied and its capacity region is determined. For the Gaussian fading models with CSIR only (transmitters have no access to CSIT), some numerical examples and simulation results are provided for Rayleigh fading.

## I. Introduction

The classical Multiple Access Channel (MAC) first considered by Shannon [1] models a communication system in which separated transmitters send independent messages to a receiver. The capacity region of the classical MAC was derived by Ahlswede [2] and Liao [3]. The MAC with cooperative encoders was first investigated by Slepian and Wolf [4], (see also [5]) wherein the capacity region of a discrete MAC with common information was established. In a MAC with common information, the transmitters send a common message (in addition to their respective private messages) to the receiver cooperatively. The concept of cooperation between transmitters in a multiple-access scenario and its effect on the capacity region has been widely studied. For a comprehensive review of the existing results for the two-transmitter/one-receive MAC with cooperative encoders, see [6] and the literature therein. In another model, the transmitters of a MAC, where a private message is associated to each one, cooperate with each other via conferencing that yields a larger capacity region. This model sometimes is referred to as MAC with partially cooperating encoders. The notion of conferencing between transmitters of MAC was first introduced by Willems [7] wherein the capacity region of a discrete memoryless two-transmitter/one-receiver MAC with conferencing encoders was established. In this scenario, encoders are connected by communication links of finite capacity, which permit both encoders to communicate over noise-free bit-pipes of given capacity. The capacity region of the Gaussian two-transmitter/one-receiver MAC with conferencing encoders was characterized in [8].

Cooperation between various nodes of a communication system via common message and conferencing was also studied for other models. Specifically, in [9] a three-transmitter/one-receiver Gaussian MAC with encoders, partially cooperating over a ring of finite-capacity unidirectional links, was studied. In [10], the capacity region of the *compound* MAC (a two-transmitter/two-receiver MAC) with a common message, also with conferencing encoders was derived. The authors in [10] also studied the interference channel with common information and also with conferencing encoders; the capacity region was established for some special cases. A two-user broadcast channel where the decoders cooperate via conferencing links was investigated in [11]. The compound MAC with a common message and conferencing decoders and also the compound MAC when both of the encoders and decoders cooperate via conferencing links was considered in [12]; the capacity region was established for some cases, specially, the physically degraded channel.

Another important feature in communication systems, specifically in mobile wireless communications, is the fading environment. The classical Gaussian fading MAC with Additive White Gaussian Noise (AWGN) was investigated in [13]-[15]. The authors in [13] and [14] characterized the capacity region of the Gaussian fading MAC where only the receiver can track the time-varying channel, i.e., the receiver has access to the Channel State Information (CSI) perfectly while the transmitters have no access to CSI. In [15] the Gaussian fading MAC with perfect CSI at the transmitters and the receiver was studied and the throughput capacities, the delay limited capacities, and the associated optimal resource allocation were obtained.

In this paper, we aim at studying the above features, i.e., cooperation among transmitters and fading environment, in a MAC, simultaneously. To this end, we consider three multiple-access scenarios and provide an exact characterization of the capacity region for each one. First, we consider a $p$-transmitter/$q$-receiver MAC coined as General MAC (GMAC) with a common message. In this network, each transmitter sends a private message over the channel as well as all transmitters send a common message cooperatively to the receivers. Each receiver decodes all transmitted messages. The capacity region of the discrete GMAC without common message was previously obtained in [16]. Here, we extend the result of [4] to the case of $p$ transmitters and $q$ receivers and characterize the capacity region of the discrete GMAC with a common message using the superposition coding technique. Then, we will consider the Gaussian fading GMAC with a common message. We show that the encoding and decoding schemes used to derive the capacity region of the discrete GMAC with a common message can constitute a frame to describe the signaling procedure for the Gaussian fading channel. In fading networks, based on the availability of CSI at different nodes (transmitters and receivers), the capacity region differs from one model to another. In a wide range of models for wireless communications, it is common to measure the state of the channel at the destinations perfectly, and then feed it to the transmitters through perfect or noisy feedback links. In our model for the Gaussian fading GMAC with a common message, we consider the channel with partial CSI at the transmitters (CSIT) and perfect CSI at the receivers (CSIR). Here, partial CSIT is modeled as the knowledge of a deterministic (potentially discrete-valued) function of CSI, where this deterministic function may vary from one transmitter to another. This model, which had been previously studied for the classical Gaussian fading MAC in [17], contains important benefits; it considers the general CSIT with full control over how much CSI is available. Specifically, it unifies the characterization of the two models in which the transmitters have perfect or no knowledge of CSI [14], [15]. Furthermore, since each transmitter is associated to its own deterministic function, the model also includes the scenario in which not all transmitters have access to CSI, which is more realistic. Under such a scenario, we prove a coding theorem for the Gaussian fading GMAC with a common message yielding an exact characterization of its throughput capacity region. It should be mentioned that the authors in [18] also studied the Gaussian fading MAC with a common message. The model studied in [18] is a special case of our derivation; it only considers a two-transmitter/one-receiver MAC ($p = 2$, and $q = 1$) with perfect CSI at both transmitters. Also, the techniques used in [18] to derive the capacity region are different from ours, yielding different formulation of the capacity region. However, we show that the result of [18] and ours in the respective special case are indeed equivalent. This yields a new proof for the result of [18].

Finally, as a third scenario, we study a two-transmitter/one-receiver Gaussian fading MAC with conferencing encoders with partial CSIT (the same model as before) and perfect CSIR, wherein it is assumed that the transmitters have access to CSIT after the conferencing. We determine the throughput capacity region of this channel.

The rest of the paper is organized as follows: In Section II, preliminaries and channel models are introduced. The main results are presented in Section III; in Subsection III-A, the GMAC with a common message and in Subsection III-B, the Gaussian fading MAC with conferencing encoders are investigated. In Section IV, numerical examples and simulation results are provided for the Gaussian Rayleigh fading. Finally, the paper is concluded in Section V.

## II. PRELIMINARIES AND DEFINITIONS

In this paper, we use the following notations: random variables (r.v.) are denoted by upper case letters (e.g., $X$) and lower case letters are used to show their realizations (e.g., $x$). The probability distribution function (p.d.f.) of a r.v. $X$ with alphabet set $\mathcal{X}$, is denoted by $P_X(x)$ where $x \in \mathcal{X}$; $P_{X|Y}(x|y)$ denotes the conditional p.d.f. of $X$ given $Y$. A sequence of r.v.'s $\langle X_t, \ldots, X_n \rangle$ with the same alphabet set $\mathcal{X}$ is denoted by $X_t^n$ and its realization is denoted by $\langle x_t, \ldots, x_n \rangle$, $x_i \in \mathcal{X}, i = t, \ldots, n$, where for the case of $X_1^n$ the subscript is dropped, occasionally. The set of all $\epsilon$-typical $n$-sequences $x^n$ with respect to the p.d.f. $P_X(x)$, as defined in [19, Ch. 3], is denoted by $A_\epsilon^n(P_X)$. The notation $\mathbb{E}[.]$ indicates the expectation operator, where sometimes to be more precise we use $\mathbb{E}_X[.]$ to denote expectation with respect to the distribution of the r.v. $X$. We also use $h(.)$ to represent the differential entropy. The set of real numbers and also the set of all nonnegative real numbers is denoted by $\mathbb{R}$ and $\mathbb{R}_+$, respectively. Finally, $\mathbb{C}(x) \equiv \frac{1}{2}\log(1+x)$.

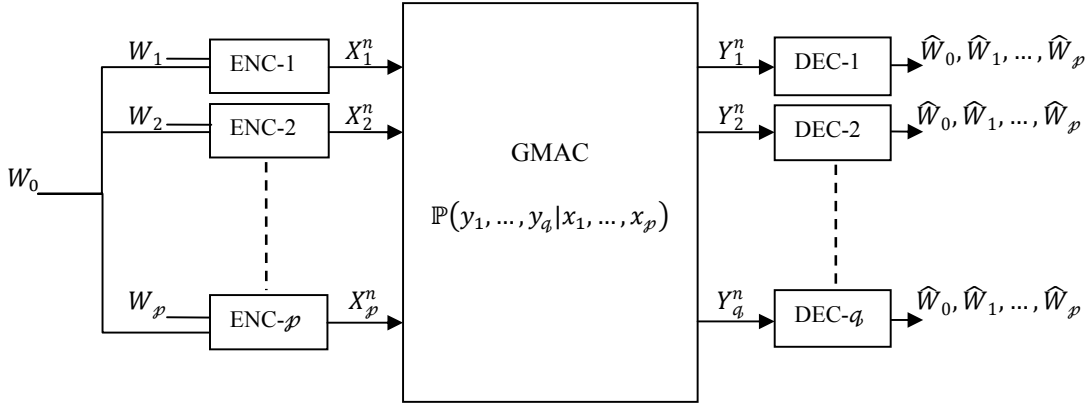

Figure 1. The GMAC with a common message.

*A) General Multiple Access Channel*

A discrete memoryless $p$-transmitter/$q$-receiver GMAC $\{\mathcal{X}_1, \mathcal{X}_2, \ldots, \mathcal{X}_p, \mathbb{P}(y_1, \ldots, y_q | x_1, \ldots, x_p), \mathcal{Y}_1, \mathcal{Y}_2, \ldots, \mathcal{Y}_q\}$, is a channel with finite input alphabets $\mathcal{X}_i, i = 1, \ldots, p$, finite output alphabets $\mathcal{Y}_j, j = 1, \ldots, q$, and a conditional p.d.f. $\mathbb{P}(y_1, \ldots, y_q | x_1, \ldots, x_p)$ that describes the relation between the inputs and outputs of the channel. The channel is assumed to be memoryless, i.e.

$$Pr(y_1^n, \ldots, y_q^n | x_1^n, \ldots, x_p^n) = \prod_{t=1}^{n} \mathbb{P}(y_{1,t}, \ldots, y_{q,t} | x_{1,t}, \ldots, x_{p,t}), \quad n \geq 1$$

(1)

where $Pr(A|B)$ is the probability of the event $A$ given $B$. Fig. 1 illustrates the channel model. In this paper, we consider a communication setting in which each transmitter sends a private message (unknown to other transmitters) over the channel, as well as all transmitters send a common message to the receivers cooperatively. Both of the common massage and the private messages are decoded at each of the receivers.

For the GMAC described by (1) with a common message, a length-$n$ code $\mathfrak{C}_n$ with a common message set $\mathcal{W}_0 = \{1, \ldots, 2^{nR_0}\}$ and private message sets $\mathcal{W}_i = \{1, \ldots, 2^{nR_i}\}, i = 1, \ldots, p$, is defined by a set of encoder functions $\{\mathfrak{E}_i(.)\}_{i=1}^{p}$ as:

$$\mathfrak{E}_i : \mathcal{W}_0 \times \mathcal{W}_i \to \mathcal{X}_i^n, \quad i = 1, \ldots, p$$

and a set of decoder functions $\{\mathfrak{D}_j(.)\}_{j=1}^{q}$ as:

$$\mathfrak{D}_j : \mathcal{Y}_j^n \to \mathcal{W}_0 \times \mathcal{W}_1 \times \ldots \times \mathcal{W}_p$$

The rate of the code is the $(\wp + 1)$-tuple $(R_0, R_1, \dots, R_\wp)$. The average error probability of decoding at the $j^{th}$ receiver $P_{e,j}^{\mathfrak{C}_n}$, $j = 1, \dots, q$, is given by:

$$P_{e,j}^{\mathfrak{C}_n} = \frac{1}{2^{n(\Sigma_{i=0}^{\wp} R_i)}} \sum_{\bar{w}_0, \bar{w}_1, \dots, \bar{w}_\wp} Pr(\{\mathfrak{D}_j(Y_j^n) \neq (\bar{w}_0, \bar{w}_1, \dots, \bar{w}_\wp)\} | (\bar{w}_0, \bar{w}_1, \dots, \bar{w}_\wp) \text{ is sent})$$

(2)

and the total average error probability of the code $P_e^{\mathfrak{C}_n}$ is given by:

$$P_e^{\mathfrak{C}_n} = \frac{1}{2^{n(\Sigma_{i=0}^{\wp} R_i)}} \sum_{\bar{w}_0, \bar{w}_1, \dots, \bar{w}_\wp} Pr\left(\left\{\bigcup_{j=1}^{n}\{\mathfrak{D}_j(Y_j^n) \neq (\bar{w}_0, \bar{w}_1, \dots, \bar{w}_\wp)\}\right\} | (\bar{w}_0, \bar{w}_1, \dots, \bar{w}_\wp) \text{ is sent}\right)$$

(3)

Note that $P_e^{\mathfrak{C}_n} \leq \sum_{j=1}^{q} P_{e,j}^{\mathfrak{C}_n}$. The $(\wp + 1)$-tuple $(R_0, R_1, \dots, R_\wp)$ of nonnegative real numbers is said to be achievable if for every $\epsilon > 0$ and for all sufficiently large $n$ there exists a length-$n$ code $\mathfrak{C}_n$ such that $P_e^{\mathfrak{C}_n} < \epsilon$. The closure of the set of all achievable rates is the capacity region.

In addition to the previous communication setting in which the input and output alphabets of the channel are of finite size, we also investigate the Gaussian fading GMAC with a common message. The channel is formulated as:

$$Y_{j,t} = \sum_{i=1}^{\wp} S_{ji,t} X_{i,t} + Z_{j,t}, \quad t \geq 1, \quad j = 1, \dots, q$$

(4)

where $\{Y_{j,t}\}_{t=1}^{\infty}, \{X_{i,t}\}_{t=1}^{\infty}$ are the $\mathbb{R}$-valued received signal at the $j^{th}$ receiver and the $\mathbb{R}$-valued transmitted signal of the $i^{th}$ transmitter, respectively; $\{S_{ji,t}\}_{t=1}^{\infty}$ is the $\mathbb{R}_+$-valued fading coefficient experienced by the signal of the $i^{th}$ transmitter at the $j^{th}$ receiver, and $\{Z_{j,t}\}_{t=1}^{\infty}$ is independent identically distributed (i.i.d.) Additive White Gaussian Noise (AWGN) process at the $j^{th}$ channel with zero mean and variance $\sigma_j^2$. The fading processes $\{S_{ji,t}\}_{t=1}^{\infty}, i = 1, \dots, \wp, j = 1, \dots, q$, are assumed to be jointly stationary and ergodic but not necessarily independent of each other. For the system given in (4) we define the *state process* of the channel as the following $q \times \wp$ matrix:

$$S_t = \begin{bmatrix} S_{11,t} & \cdots & S_{1\wp,t} \\ \vdots & \ddots & \vdots \\ S_{q1,t} & \cdots & S_{q\wp,t} \end{bmatrix}_{q \times \wp}, \quad t \geq 1$$

(5)

Furthermore, we assume that the state process of the channel $\{S_t\}_{t=1}^{\infty}$ is independent of the AWGN process $\{Z_{j,t}\}_{t=1}^{\infty}, j = 1, \dots, q$.

*Partial CSIT and perfect CSIR:* In this paper, we consider the channel model in (4) with partial CSIT; the $i^{th}$ transmitter at each time instant knows a version of the current CSI determined by a deterministic function of the state process of the channel. In other words, the CSI available at the $i^{th}$ transmitter is given by $\xi_i: [\mathbb{R}_+]_{q \times \wp} \to \mathcal{E}_i, i = 1, \dots, \wp$, where $[\mathbb{R}_+]_{q \times \wp}$ is the space of all $q \times \wp$ matrices with $\mathbb{R}_+$-valued entries and $\mathcal{E}_i$ is an arbitrary set (maybe finite) associated to the $i^{th}$ transmitter. Thus, for the channel in (4) the $i^{th}$ encoder consists of a set of functions $\{\mathfrak{E}_{i,t}\}_{t=1}^{n}$ where:

$$\mathfrak{E}_{i,t}: \mathcal{W}_0 \times \mathcal{W}_i \times \mathcal{E}_i \to \mathbb{R}, \quad t = 1, \dots, n, \quad i = 1, \dots, \wp$$

and the input symbol of the channel due to the $i^{th}$ transmitter at each time instant $t$, is given by $X_{i,t} = \mathfrak{E}_{i,t}(W_0, W_i, E_{i,t})$, where $E_{i,t} = \xi_i(S_t) \in \mathcal{E}_i$. Furthermore, we assume that the receivers have access to perfect CSI, i.e., the $j^{th}$ decoder function $j = 1, \dots, q$, is given by:

$$\mathfrak{D}_j: \mathcal{Y}_j^n \times \left([\mathbb{R}_+]_{q \times p}\right)^n \to \mathcal{W}_0 \times \mathcal{W}_1 \times \ldots \times \mathcal{W}_p$$

which estimates the messages as:

$$\left(\widehat{W}_0^{(j)}, \widehat{W}_1^{(j)}, \ldots, \widehat{W}_p^{(j)}\right) = \mathfrak{D}_j(Y_j^n, S^n)$$

where $S^n = (S_1, S_2, \ldots, S_n)$ is the state process of the channel.

An average power constraint is imposed on the codewords of each encoder. Thus, the $i^{th}$ transmitter is subjected to an average power constraint $P_i \in \mathbb{R}_+$, in the following way:

$$\frac{1}{n}\mathbb{E}\left[\sum_{t=1}^{n}\left(X_{i,t}(W_0, W_i, \xi_i(S_t))\right)^2\right] \leq P_i, \quad i = 1, \ldots, p.$$

(6)

*B) Multiple Access Channels with Conferencing Encoders*

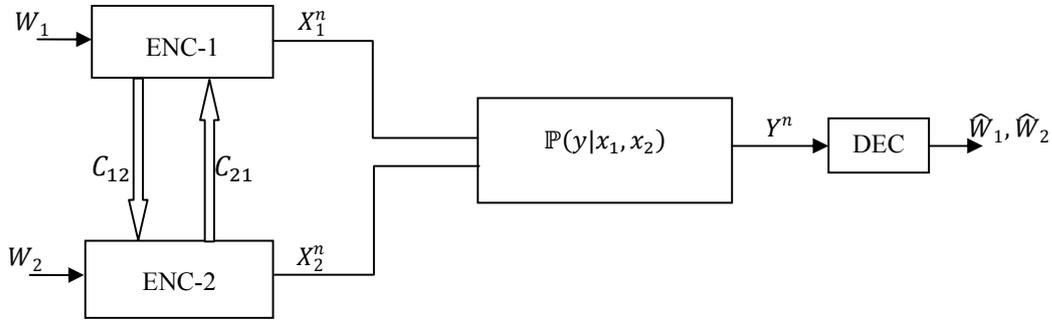

Figure 2. The MAC with conferencing encoders.

A discrete two-transmitter/one-receiver MAC with conferencing encoders $\{\mathcal{X}_1, \mathcal{X}_2, \mathbb{P}(y|x_1, x_2), \mathcal{Y}, C_{12}, C_{21}\}$, is a channel with two (finite) input alphabet sets $\mathcal{X}_1, \mathcal{X}_2$, an (finite) output alphabet set and a conditional probability distribution $\mathbb{P}(y|x_1, x_2)$ which characterizes the channel. In addition, encoders are connected to each other with two links of capacities $C_{12}, C_{21}$. Fig. 2 depicts the channel model.

A length-$n$ code $\mathfrak{C}(n, K, C_{12}, C_{21})$ for the MAC with conferencing encoders has two message sets $\mathcal{W}_i = \{1, \ldots, 2^{nR_i}\}, i = 1,2$: before transmitting of messages over the channel, the two encoders hold a conference, i.e., the code $\mathfrak{C}(n, K, C_{12}, C_{21})$ consists of two sets of $K$ communicating functions $\{\hbar_{i,1}, \ldots, \hbar_{i,K}\}, i = 1,2$, and two sets of (finite) conferencing alphabets $\{\mathcal{V}_{i,1}, \ldots, \mathcal{V}_{i,K}\}, i = 1,2$. Each communicating function $\hbar_{i,k}, i = 1,2$, for $k = 1, \ldots, K$, maps the message $W_i$ and the sequence of the previously received symbols from the other transmitter into the $k^{th}$ symbol $V_{i,k} \in \mathcal{V}_{i,k}$. In notations we have:

$$\hbar_{1,k}: \mathcal{W}_1 \times \mathcal{V}_2^{k-1} \to \mathcal{V}_{1,k}, \quad V_{1,k} = \hbar_{1,k}(W_1, V_2^{k-1}),$$
$$\hbar_{2,k}: \mathcal{W}_2 \times \mathcal{V}_1^{k-1} \to \mathcal{V}_{2,k}, \quad V_{2,k} = \hbar_{2,k}(W_2, V_1^{k-1}),$$

(7)

A conference is said to be $(C_{12}, C_{21})$-permissible [7] if the sets of communicating functions are such that:

$$\sum_{k=1}^{K} \log\|\mathcal{V}_{1,k}\| \leq nC_{12}, \quad \sum_{k=1}^{K} \log\|\mathcal{V}_{2,k}\| \leq nC_{21}$$

(8)

where $\|\mathcal{V}_{i,k}\|$ denotes the cardinality of $\mathcal{V}_{i,k}$.

After the conferencing, transmitter 1 knows the sequence $V_2^K = (V_{2,1}, \ldots, V_{2,K})$ and transmitter 2 knows the sequence $V_1^K = (V_{1,1}, \ldots, V_{1,K})$. Furthermore, the code $\mathfrak{C}(n, K, C_{12}, C_{21})$ has two encoder functions $\{\mathfrak{E}_1, \mathfrak{E}_2\}$ described as:

$$\mathfrak{E}_1: \mathcal{W}_1 \times \mathcal{V}_2^K \to \mathcal{X}_1^n, \qquad \mathfrak{E}_2: \mathcal{W}_2 \times \mathcal{V}_1^K \to \mathcal{X}_2^n$$

also the decoder is given by:

$$\mathfrak{D}: \mathcal{Y}^n \to \mathcal{W}_1 \times \mathcal{W}_2.$$

The rate of the code $\mathfrak{C}(n, K, C_{12}, C_{21})$ is the pair $(R_1, R_2)$ and the error probability of the code $P_e^{\mathfrak{C}(n,K,C_{12},C_{21})}$ is defined as:

$$P_e^{\mathfrak{C}(n,K,C_{12},C_{21})} = \frac{1}{2^{n(R_1+R_2)}} \sum_{\overline{w}_1, \overline{w}_2} Pr(\{\mathfrak{D}(Y^n) \neq (\overline{w}_1, \overline{w}_2)\} | (\overline{w}_1, \overline{w}_2) \text{ is sent}).$$

(9)

A rate pair $(R_1, R_2)$ of nonnegative numbers is said to be achievable for the MAC with conferencing encoders if for every $\epsilon > 0$ and for all sufficiently large $n$, there exists a code $\mathfrak{C}(n, K, C_{12}, C_{21})$ such that the resulting conference of the code is $(C_{12}, C_{21})$-permissible and $P_e^{\mathfrak{C}(n,K,C_{12},C_{21})} < \epsilon$. The capacity region of the MAC with conferencing encoders is the closure of all achievable rate pairs.

The MAC with conferencing encoders was first introduced in [7] where the capacity region of the discrete channel model was derived. In this paper, we investigate the Gaussian fading MAC with conferencing encoders in which the channel is formulated as:

$$Y_t = S_{1,t} X_{1,t} + S_{2,t} X_{2,t} + Z_t, \qquad t \geq 1$$

(10)

where $\{Y_t\}_{t=1}^{\infty}, \{X_{i,t}\}_{t=1}^{\infty}, i = 1,2$, are the $\mathbb{R}$-valued received signal and the $\mathbb{R}$-valued transmitted signal of the $i^{th}$ transmitter, respectively; $\{S_{i,t}\}_{t=1}^{\infty}, i = 1,2$, is the $\mathbb{R}_+$-valued fading coefficient experienced by the signal of the $i^{th}$ transmitter, and $\{Z_t\}_{t=1}^{\infty}$ is i.i.d. AWGN process with zero mean and variance $\sigma^2$. The fading processes $\{S_{i,t}\}_{t=1}^{\infty}, i = 1,2$, are assumed to be jointly stationary and ergodic but not necessarily independent of each other. For the channel given by (10) we define the state process of the channel as $S_t = (S_{1,t}, S_{2,t}), t \geq 1$. The definition of conference for the channel in (10) is the same as discrete channel and is given by (7) and (8). Similar to the communication system in (4), we consider the Gaussian fading MAC with conferencing encoders with partial CSIT and perfect CSIR, where CSIT is known to the encoders after the conferencing. At each time instant, CSIT at the $i^{th}$ transmitter, $i = 1,2$, is a deterministic function of the current state of channel. Let $\xi_i: \mathbb{R}_+^2 \to \mathcal{E}_i$ be the mapping that describes CSIT at the $i^{th}$ transmitter, where $\mathcal{E}_i$ is an arbitrary set (maybe finite) associated to the $i^{th}$ transmitter, $i = 1,2$. Assume that the information that transmitter 1 obtains after the conferencing is $V_2^K = (V_{2,1}, \ldots, V_{2,K})$ and for transmitter 2 is $V_1^K = (V_{1,1}, \ldots, V_{1,K})$. By these assumptions the $i^{th}$ encoder consists of a set of functions $\{\mathfrak{E}_{i,t}\}_{t=1}^n, i = 1,2$, where for $t = 1, \ldots, n$:

$$\begin{cases} \mathfrak{E}_{1,t}: \mathcal{W}_1 \times \mathcal{V}_2^K \times \mathcal{E}_1 \to \mathbb{R}, \\ \mathfrak{E}_{2,t}: \mathcal{W}_2 \times \mathcal{V}_1^K \times \mathcal{E}_2 \to \mathbb{R} \end{cases}$$

Thus, the input symbol of the channel due to transmitter 1 at each time instant $t$ is given by $X_{1,t} = \mathfrak{E}_{1,t}(W_1, V_2^n, E_{1,t})$, where $E_{1,t} = \xi_1(S_t) \in \mathcal{E}_1$, (respectively for transmitter 2, $X_{2,t} = \mathfrak{E}_{2,t}(W_2, V_1^n, E_{2,t})$, where $E_{2,t} = \xi_2(S_t) \in \mathcal{E}_2$). We also assume that the receiver knows perfect CSI, i.e., the decoder function is given by:

$$\mathfrak{D}: \mathcal{Y}^n \times (\mathbb{R}_+^2)^n \to \mathcal{W}_1 \times \mathcal{W}_2$$

which estimates the messages as:

$$(\widehat{W}_1, \widehat{W}_2) = \mathfrak{D}(Y^n, S^n)$$

where $S^n = (S_1, S_2, \ldots, S_n)$ is the state process of the channel.

Finally, we note that an average power constraint is also imposed on the codewords of each encoder and the $i^{th}$ transmitter is subjected to an average power constraint $P_i \in \mathbb{R}_+$, $i = 1,2$. Precisely speaking, for the codewords of transmitter 1:

$$\frac{1}{n}\mathbb{E}\left[\sum_{t=1}^{n}\left(X_{1,t}(W_1, V_2^n, \xi_1(S_t))\right)^2\right] \leq P_1$$

(11)

and to obtain the power constraint on the codewords of transmitter 2 one can exchange the indices 1 and 2 in (11).

In the next section, we state our main results for the GMAC with a common message and also for the Gaussian fading MAC with conferencing encoders.

### III. MAIN RESULTS

*III-A) General multiple access channel with a common message*

The main result of this subsection is given in Theorem 1 where we provide an exact characterization of the capacity region for the Gaussian fading GMAC defined by (4) with common message. Firstly, we derive the capacity region of the discrete memoryless GMAC (1) with a common message using the superposition coding technique. As we see later, the achievability proof of the capacity region for this channel is useful to obtain the capacity region for the Gaussian fading GMAC (4) with common message.

*Proposition 1*: The capacity region of the discrete memoryless GMAC with $p$ transmitters and $q$ receivers and with common information denoted by $\mathcal{C}_{cm}^{D-GMAC}$, is given by:

$$\mathcal{C}_{cm}^{D-GMAC} = \bigcup \left\{\begin{array}{l}(R_0, R_1, R_2, \ldots, R_p) \in \mathbb{R}_+^{p+1}\\ \forall \Lambda \subseteq \{1, \ldots, p\}, \forall j \in \{1, \ldots, q\}\\ \sum_{k \in \Lambda} R_k \leq I(\{X_k\}_{k \in \Lambda}; Y_j | \{X_k\}_{k \notin \Lambda}, U)\\ ,\\ \sum_{k=0}^{p} R_k \leq I(X_1, \ldots, X_p; Y_j)\end{array}\right\}$$

(12)

where $U$ is an auxiliary random variable and the joint p.d.f. of the r.v.'s $\langle U, X_1, \ldots, X_p, Y_1, \ldots, Y_q \rangle$ is given by:

$$P_{UX_1 \ldots X_p Y_1 \ldots Y_q}(u, x_1, \ldots, x_p, y_1, \ldots, y_q) = P_U(u) \prod_{i=1}^{p} P_{X_i|U}(x_i|u) \, \mathbb{P}(y_1, \ldots, y_q | x_1, \ldots, x_p)$$

(13)

*Proof of Proposition 1:* This proposition is an extension of the result of [4] to the case of $p$ transmitters and $q$ receivers. In Appendix I, we provide a complete proof of the direct part (achievability proof); similar encoding and decoding schemes are exploited to obtain the capacity region of the Gaussian fading GMAC (4) with common message, as shown in the next theorem. The proof of the converse part will be omitted for brevity.

*Remarks:*

1) The cardinality of the auxiliary r.v. $U$ is bounded above as $\|\mathcal{U}\| \leq \prod_{i=1}^{p}\|\mathcal{X}_i\| + q \times 2^p - 1$. This can be proved through the standard methods using the Support lemma [20, p. 310].
2) By setting $p = 2$ and $q = 1$, the result of Proposition 1 reduces to the capacity region of a two user discrete MAC with a common message obtained in [4].
3) By setting $p = q = 2$, the result of Proposition 1 reduces to the capacity region of a compound MAC with a common message derived in [10].

4) By setting $U \equiv \emptyset$ and $R_0 = 0$, the result of Proposition 1 reduces to the capacity region of a GMAC without common message derived in [16].

Now, we consider the Gaussian fading GMAC (4) with common message. In the following theorem, we characterize the throughput capacity region [15] of this channel.

*Theorem 1:* Consider the Gaussian fading GMAC in (4) with a common message, with partial CSIT and perfect CSIR. The capacity region denoted by $C_{cm}^{Gf-GMAC}$, is given by:

$$C_{cm}^{Gf-GMAC} = \bigcup_{\substack{\{\varphi_i(.): \mathbb{E}[\varphi_i(E_i)] \leq P_i\}_{i=1}^{\wp} \\ \{\varrho_i(.): 0 \leq \varrho_i(.) \leq 1\}_{i=1}^{\wp}}} \left\{ \begin{array}{l} (R_0, R_1, R_2, \ldots, R_\wp) \in \mathbb{R}_+^{\wp+1} \\ \forall \Lambda \subseteq \{1, \ldots, \wp\}, \forall j \in \{1, \ldots, q\}: \\ \sum_{k \in \Lambda} R_k \leq \mathbb{E}_S \left[ \mathbb{C} \left( \frac{\sum_{k \in \Lambda} \left( (S_{jk})^2 \varphi_k(E_k) \left(1 - \varrho_k^2(E_k)\right) \right)}{\sigma_j^2} \right) \right] \\ , \\ \sum_{k=0}^{\wp} R_k \leq \mathbb{E}_S \left[ \mathbb{C} \left( \frac{\sum_{k=1}^{\wp} (S_{jk})^2 \varphi_k(E_k)}{\sigma_j^2} \right) \right] \end{array} \right\}$$

(14)

where $S \in [\mathbb{R}_+]_{q \times \wp}$ is a r.v. representing the state process of the channel (the matrix of fading coefficients) and $S_{ij} \in \mathbb{R}_+, i = 1, \ldots, \wp, j = 1, \ldots, q$, is a r.v. representing the corresponding entry of $S$, $E_i \in \mathcal{E}_i, i = 1, \ldots, \wp$, is a r.v. representing the CSIT available at the $i^{th}$ transmitter which is a deterministic function of $S$: $E_i = \xi_i(S)$. In addition, the function $\varphi_i: \mathcal{E}_i \to \mathbb{R}_+$ satisfying the following constraint:

$$\mathbb{E}[\varphi_i(E_i)] \leq P_i, \quad i = 1, \ldots, \wp$$

(15)

denotes the power allocation policy for the $i^{th}$ transmitter and $\varrho_i: \mathcal{E}_i \to [0,1], i = 1, \ldots, \wp$, is an arbitrary (limited) deterministic function that takes its values from the interval $[0,1]$.

*Remarks:*

1) By setting $q = 1$, and $R_0 = 0$, $\varrho_k(.) \equiv 0, k = 1, \ldots, \wp$, and also $E_k \equiv S, k = 1, \ldots, \wp$, (perfect CSIT), then the rate region (14) reduces to the capacity region of a $\wp$-user Gaussian fading MAC without common message with perfect CSIT and also perfect CSIR, obtained previously in [15, Part I, Th. 2.1].
2) We observe that the throughput capacity region of the Gaussian fading GMAC with common message given in (14) depends only on the stationary (first order) distribution of the joint fading processes, i.e., $P_S(s)$ where $S$ is the matrix of fading coefficients and not on the correlation structure (memory of the state process). A same observation could be found in [15, Part I, Th. 2.1] for the capacity of Gaussian fading MAC without common message, as indicated in [15, Part II].
3) As we see from (14), the improvement in the capacity because of having partial knowledge of the CSI at the transmitters comes from the ability to allocate powers (according to the functions $\varphi_i: \mathcal{E}_i \to \mathbb{R}^+, i = 1, \ldots, \wp$, with constraints (15)) and also from the functions $\varrho_i: \mathcal{E}_i \to [0,1], i = 1, \ldots, \wp$. The functions $\varrho_i(.)$ appeared in the capacity formulation (14) arise from the existence of the common message in the system which causes that the transmitted signals to be correlated. Precisely speaking, by (24) the transmitted signals of $i^{th}$ and $j^{th}$ encoders are correlated according to a correlation coefficient denoted by $Y_{i,j}, i, j = 1, \ldots, \wp$, which is given as follows:

$$Y_{i,j} \triangleq \frac{\mathbb{E}[X_i X_j]}{\sqrt{\mathbb{E}[X_i^2]\mathbb{E}[X_j^2]}} = \frac{\mathbb{E}\left[\sqrt{\varphi_i(E_i)\varphi_j(E_j)}\varrho_i(E_i)\varrho_j(E_j)\right]}{\sqrt{\mathbb{E}[\varphi_i(E_i)]\mathbb{E}[\varphi_j(E_j)]}}$$

(16)

As mentioned before, for the case where there is no common message in the system these functions disappear, i.e., $\varrho_k(.) \equiv 0, k = 1, \ldots, \wp$; therefore, the improvement in the capacity comes only from the ability to allocate powers [15, Part II, Th. 2.1] and the transmitted signals are uncorrelated.

4) *Lagrangian characterization of the capacity region:* One can solve for the boundary surface of the rate region (14) to explicitly characterize the capacity region. Due to the convexity of the capacity region the boundary surface of $\mathcal{C}_{cm}^{Gf-GMAC}$ given by (14), is the closure of all points $\boldsymbol{R}^* = (R_0^*, R_1^*, R_2^*, \ldots, R_\wp^*)$ such that $\boldsymbol{R}^*$ is a solution to the following optimization problem:

$$\max_{(R_0, R_1, R_2, \ldots, R_\wp)} \sum_{i=0}^{\wp} \mu_i R_i, \quad \text{subject to } (R_0, R_1, R_2, \ldots, R_\wp) \in \mathcal{C}_{cm}^{Gf-GMAC} \tag{17}$$

For some $(\mu_0, \mu_1, \mu_2, \ldots, \mu_\wp) \in \mathbb{R}_+^{\wp+1}$. Now consider the following set:

$$\partial \triangleq \{(\boldsymbol{R}, \boldsymbol{P}) : \boldsymbol{P} \in \mathbb{R}_+^{\wp}, \boldsymbol{R} \in \mathcal{C}_{cm}^{Gf-GMAC}(\boldsymbol{P})\} \tag{18}$$

where $\mathcal{C}_{cm}^{Gf-GMAC}(\boldsymbol{P})$ represents the capacity assuming that the power constraint associated to the $i^{th}$ transmitter is given by the $i^{th}$ entry of $\boldsymbol{P}$. By concavity of the logarithm function, it can be shown that $\partial$ is a convex set. Therefore, there exist Lagrange multipliers $(\lambda_1, \lambda_2, \ldots, \lambda_\wp) \in \mathbb{R}_+^{\wp}$, such that $\boldsymbol{R}^*$ is a solution to the following equivalent optimization problem:

$$\max_{(\boldsymbol{R}, \boldsymbol{P}) \in \partial} \sum_{i=0}^{\wp} \mu_i R_i - \sum_{i=1}^{\wp} \lambda_i P_i \tag{19}$$

As a fact, the capacity region (14) remains unchanged if we replace the power constraint inequalities in (15) with equality. Furthermore, the $\mathcal{C}_{cm}^{Gf-GMAC}$ given by (14) is a union taken over the set of all functions $\{\varphi_i(.) : \mathbb{E}[\varphi_i(E_i)] = P_i\}_{i=1}^{\wp}$ and $\{\varrho_i(.) : 0 \leq \varrho_i(.) \leq 1\}_{i=1}^{\wp}$. Thus we can rewrite (19) as an optimization over the set of such functions, as follows:

$$\max_{\substack{(\boldsymbol{R}, \{\varphi_i(.): 0 \leq \varphi_i(.)\}_{i=1}^{\wp}, \\ \{\varrho_i(.): 0 \leq \varrho_i(.) \leq 1\}_{i=1}^{\wp})}} \sum_{i=0}^{\wp} \mu_i R_i - \sum_{i=1}^{\wp} \lambda_i \mathbb{E}[\varphi_i(E_i)],$$

$$\text{subject to } \boldsymbol{R} \in \mathcal{C}_{cm}^{Gf-GMAC}\big(\{\varphi_i(.)\}_{i=1}^{\wp}, \{\varrho_i(.): 0 \leq \varrho_i(.) \leq 1\}_{i=1}^{\wp}\big) \tag{20}$$

where $\mathcal{C}_{cm}^{Gf-GMAC}\big(\{\varphi_i(.)\}_{i=1}^{\wp}, \{\varrho_i(.): 0 \leq \varrho_i(.) \leq 1\}_{i=1}^{\wp}\big)$ is the capacity where we fix $\varphi_i(.)$-functions and $\varrho_i(.)$-functions to be as in the arguments. Solving this optimization problem is not the subject of this paper and we relegate it to future work.

*Proof of Theorem 1:* We prove this theorem in two parts; first we prove the achievability of (14) through a random coding technique and then we provide a converse theorem.

<u>Achievability part</u>: It is worth noting that for the case where there is no common message in the system, the optimal signaling for the channel is rather trivial (see [15, App. A, p. 2812]). However, the same does not hold for the channel with common message. Here, we prove the achievability scheme through a random coding argument where the encoding and decoding procedure is based on the superposition technique. Let $\langle U, V_1, \ldots, V_\wp \rangle$ be a sequence of auxiliary r.v.'s with alphabets $U \in \mathcal{U}$ and $V_i \in \mathcal{V}_i, i = 1, \ldots, \wp$. First, we show that for the Gaussian fading GMAC (4) with a common message, with partial CSIT and perfect CSIR, the following rate region is achievable:

$$\bigcup_{\substack{P_U \prod_{i=1}^{p} P_{V_i|U}, \\ \{f_i(.): \mathcal{V}_i \times \mathcal{U} \times \mathcal{E}_i \to \mathbb{R}\}_{i=1}^{p}: \\ X_i = f_i(V_i, U, E_i)}} \left\{ \begin{array}{l} (R_0, R_1, R_2, \ldots, R_p) \in \mathbb{R}_+^{p+1} \\ \forall \Lambda \subseteq \{1, \ldots, p\}, \forall j \in \{1, \ldots, q\} \\ \sum_{k \in \Lambda} R_k \leq I(\{V_k\}_{k \in \Lambda}; Y_j | \{V_k\}_{k \notin \Lambda}, U, S) \\ \sum_{k=0}^{p} R_k \leq I(U, V_1, \ldots, V_p; Y_j | S) \end{array} \right\}$$

(21)

where:

$$Y_j = \sum_{k=1}^{p} S_{jk} X_k + Z_j, \qquad j = 1, \ldots, q,$$

(22)

and the joint p.d.f. of the r.v.'s $\langle S, E_1, \ldots, E_p, U, V_1, \ldots, V_p \rangle$ is given by:

$$P_{SE_1 \ldots E_p U V_1 \ldots V_p}(s, e_1, \ldots, e_p, u, v_1, \ldots, v_p) = P_{SE_1 \ldots E_p}(s, e_1, \ldots, e_p) P_U(u) \prod_{i=1}^{p} P_{V_i|U}(v_i|u).$$

(23)

Furthermore, $\{f_i(.): \mathcal{V}_i \times \mathcal{U} \times \mathcal{E}_i \to \mathbb{R}\}_{i=1}^{p}$ is a set of deterministic functions such that $X_i = f_i(V_i, U, E_i), i = 1, \ldots, p$, satisfies the power constraint policy of the $i^{th}$ transmitter: $\mathbb{E}[X_i^2] \leq P_i$.

To prove this, we use a random codebook generation technique as follows:

1) Fix the distribution $P_{UV_1 \ldots V_p}(u, v_1, \ldots, v_p) = P_U(u) \prod_{i=1}^{p} P_{V_i|U}(v_i|u)$ and the set of deterministic functions $\{f_i(.): \mathcal{V}_i \times \mathcal{U} \times \mathcal{E}_i \to \mathbb{R}\}_{i=1}^{p}$ such that $\mathbb{E}\left[\left(f_i(V_i, U, E_i)\right)^2\right] \leq P_i, i = 1, \ldots, p$.
2) Generate at random $2^{nR_0}$ i.i.d. sequences $u^n$ according to $P(u^n) = \prod_{t=1}^{n} P_U(u_t)$. Label these sequences $u^n(w_0)$, where $w_0 \in \{1, \ldots, 2^{nR_0}\}$.
3) For each $u^n(w_0)$, generate $2^{nR_i}$, $i = \{1, 2, \ldots, p\}$, sequences $v_i^n$ according to the p.d.f. $P(v_i^n | u^n(w_0)) = \prod_{t=1}^{n} P_{V_i|U}(v_{i,t} | u_t)$. Label these sequences $v_i^n(w_0, w_i)$, where $w_i \in \{1, \ldots, 2^{nR_i}\}$.

*Encoding:* To send a common message $w_0$ and a private message $w_i$, at each time instant $t, 1 \leq t \leq n$, the $i^{th}$ encoder assuming observation of $e_{i,t} \in \mathcal{E}_i$ as CSIT, sends $x_{i,t} = f_i(v_{i,t}(w_0, w_i), u_t(w_0), e_{i,t})$ over the channel.

*Decoding at the $j^{th}$ receiver, $j \in \{1, 2, \ldots, q\}$:* The $j^{th}$ receiver assuming reception of the signal sequence $y_j^n$ and tracking the state process of the channel perfectly as $s^n$, tries to find a unique $(\widehat{w}_0, \widehat{w}_1, \ldots, \widehat{w}_p)$ such that $\left(u^n(\widehat{w}_0), v_1^n(\widehat{w}_0, \widehat{w}_1), \ldots, v_p^n(\widehat{w}_0, \widehat{w}_p), y_j^n, s^n\right) \in A_\epsilon^n\left(P_{SUV_1 \ldots V_p Y_j}\right)$.

By analysis of the error probability that is similar to the proof of direct part of Proposition 1 (given in Appendix I), and noting that the state process is stationary, the channel is memoryless and the codewords $\langle U^n, V_1^n, \ldots, V_p^n \rangle$ were generated i.i.d. and independent of the state process $S^n$, one can see that the probability of error tends to zero provided that the rate tuple $(R_0, R_1, R_2, \ldots, R_p)$ belongs to the rate region (21).

Now, to prove the achievability of (14) let $\langle U, V_1, V_2, \ldots, V_p \rangle$ be a sequence of Gaussian-distributed r.v.'s with zero mean and unit variance, independent of each other and also independent of the state matrix $S$;

In addition, let $\{\varphi_i: \mathcal{E}_i \to \mathbb{R}_+\}_{i=1}^{p}$ be a set of power allocation policy functions which satisfy the power constraints in (15) and $\{\varrho_i: \mathcal{E}_i \to [0,1]\}_{i=1}^{p}$ be a set of arbitrary deterministic functions with range $[0,1]$. Define the r.v.'s $\langle X_1, X_2, \ldots, X_p \rangle$ as:

$$X_i \triangleq \sqrt{\varphi_i(E_i)}\left(\varrho_i(E_i)U + \sqrt{1-\varrho_i^2(E_i)}V_i\right), \qquad i=1,\ldots,p.$$

(24)

Note that in definition (24), $\langle E_1,\ldots,E_p\rangle$ is the sequence of CSITs at the transmitters with joint p.d.f. determined by $P_{SE_1\ldots E_p}(s,e_1,\ldots,e_p)$. One can easily check that the r.v.'s $\langle X_1, X_2,\ldots,X_p\rangle$ defined in (24) satisfy:

$$\mathbb{E}_S[X_i^2] = \mathbb{E}_S[\varphi_i(E_i)] \leq P_i, \qquad i=1,\ldots,p.$$

(25)

Thus by substituting $\langle U, V_1, V_2,\ldots,V_p\rangle$ and $\langle X_1, X_2,\ldots,X_p\rangle$ as defined in (24) in the rate region given by (21) we have:

$\forall \Lambda \subseteq \{1,\ldots,p\}, \forall j \in \{1,\ldots,q\}$:

$$I(\{V_k\}_{k\in\Lambda}; Y_j | \{V_k\}_{k\notin\Lambda}, U, S) \stackrel{(a)}{=} h(Y_j | \{V_k\}_{k\notin\Lambda}, \{X_k\}_{k\notin\Lambda}, U, S) - h(Y_j | \{V_k\}_{k=1}^p, \{X_k\}_{k=1}^p, U, S)$$

$$= h\left(\sum_{k=1}^p S_{jk}X_k + Z_j \middle| \{V_k\}_{k\notin\Lambda}, \{X_k\}_{k\notin\Lambda}, U, S\right) - h\left(\sum_{k=1}^p S_{jk}X_k + Z_j \middle| \{V_k\}_{k=1}^p, \{X_k\}_{k=1}^p, U, S\right)$$

$$\stackrel{(b)}{=} h\left(\sum_{k\in\Lambda} S_{jk}X_k + Z_j \middle| U, S\right) - \frac{1}{2}\log(2\pi e\sigma_j^2)$$

$$= h\left(\sum_{k\in\Lambda} S_{jk}\sqrt{\varphi_k(E_k)}\left(\varrho_k(E_k)U + \sqrt{1-\varrho_k^2(E_k)}V_k\right) + Z_j \middle| U, S\right) - \frac{1}{2}\log(2\pi e\sigma_j^2)$$

$$= h\left(\sum_{k\in\Lambda} S_{jk}\sqrt{\varphi_k(E_k)}\left(\sqrt{1-\varrho_k^2(E_k)}V_k\right)V_k + Z_j \middle| S\right) - \frac{1}{2}\log(2\pi e\sigma_j^2)$$

$$= \mathbb{E}_S\left[h\left(\sum_{k\in\Lambda} s_{jk}\sqrt{\varphi_k(\xi_k(s))}\left(\sqrt{1-\varrho_k^2(\xi_k(s))}V_k\right) + Z_j \middle| S=s\right)\right] - \frac{1}{2}\log(2\pi e\sigma_j^2)$$

$$= \mathbb{E}_S\left[\mathbb{C}\left(\frac{\sum_{k\in\Lambda}\left((s_{jk})^2 \varphi_k(E_k)\left(1-\varrho_k^2(E_k)\right)\right)}{\sigma_j^2}\right)\right]$$

(26)

where the equality (a) is due to the fact that $X_k$ is known given $(V_k, U, E_k = \xi_k(S))$ and (b) is because from (24) $(\{V_k\}_{k\notin\Lambda}, \{X_k\}_{k\notin\Lambda}) \to U, S \to \{X_k\}_{k\in\Lambda}$ forms a Markov chain. Furthermore, we can write:

$$I(U, V_1,\ldots,V_p; Y_j | S) = h\left(\sum_{k=1}^p S_{jk}X_k + Z_j \middle| S\right) - h\left(\sum_{k=1}^p S_{jk}X_k + Z_j \middle| U, \{V_k\}_{k=1}^p, \{X_k\}_{k=1}^p, S\right)$$

$$= h\left(\sum_{k=1}^p S_{jk}\sqrt{\varphi_k(E_k)}\left(\varrho_k(E_k)U + \sqrt{1-\varrho_k^2(E_k)}V_k\right) + Z_j \middle| S\right) - \frac{1}{2}\log(2\pi e\sigma_j^2)$$

$$= \mathbb{E}_S\left[h\left(\sum_{k=1}^p s_{jk}\sqrt{\varphi_k(\xi_k(s))}\left(\varrho_k(\xi_k(s))U + \sqrt{1-\varrho_k^2(\xi_k(s))}V_k\right) + Z_j \middle| S=s\right)\right] - \frac{1}{2}\log(2\pi e\sigma_j^2)$$

$$= \mathbb{E}_S \left[ \mathbb{C} \left( \frac{\sum_{k=1}^{p}(S_{jk})^2 \varphi_k(E_k) + 2\sum_{1\leq k<i\leq p}(S_{jk}S_{ji}\varrho_k(E_k)\varrho_i(E_i)\sqrt{\varphi_k(E_k)\varphi_i(E_i)})}{\sigma_j^2} \right) \right]$$

(27)

Now, by substituting (26) and (27) in the rate region (21) we obtain the achievability of (14).

*Converse part*: To prove the converse part we first derive an outer bound on $\mathcal{C}_{cm}^{Gf-GMAC}$ in terms of mutual information functions, in the following lemma.

*Lemma 1*: $\mathcal{C}_{cm}^{Gf-GMAC}$ is outer bounded by:

$$\mathcal{C}_{cm}^{Gf-GMAC} \subseteq \bigcup \left\{ \begin{array}{l} (R_0, R_1, R_2, \ldots, R_p) \in \mathbb{R}_+^{p+1} \\ \forall \Lambda \subseteq \{1, \ldots, p\}, \forall j \in \{1, \ldots, q\}, \\ \sum_{k \in \Lambda} R_k \leq I(\{X_k\}_{k \in \Lambda}; Y_j | \{X_k\}_{k \notin \Lambda}, U, S) \\ , \\ \sum_{k=0}^{p} R_k \leq I(X_1, \ldots, X_p; Y_j | S) \end{array} \right\}$$

(28)

over all joint p.d.f. of the form:

$$P_{SE_1\ldots E_p UX_1\ldots X_p}(s, e_1, \ldots, e_p, u, x_1, \ldots, x_p) = P_{SE_1\ldots E_p}(s, e_1, \ldots, e_p) P_U(u) \prod_{i=1}^{p} P_{X_i|UE_i}(x_i|u, e_i)$$

(29)

that satisfy the power constraints: $\mathbb{E}[X_i^2] \leq P_i, i = 1, \ldots, p$; where $S \in [\mathbb{R}_+]_{q \times p}$ is a r.v. representing the state process of the channel defined in (5) and $E_i \in \mathcal{E}_i$ is the r.v. representing the CSIT available at the $i^{th}$ transmitter that is (by definition of the channel) a deterministic function of $S$: $E_i = \xi_i(S), i = 1, \ldots, p$. We also note that $P_{SE_1\ldots E_p}(s, e_1, \ldots, e_p)$ is obtained from the (first order) distribution of the channel state process, i.e., $P_S(s)$, which describes the behavior of the (matrix) fading coefficients.

*Proof*: See Appendix II.

Now, to see that the rate region (14) is also an outer bound for the capacity region of the underlying channel, i.e., $\mathcal{C}_{cm}^{Gf-GMAC}$, it is sufficient to show that is a superset of (28). However, we prove a stronger result that is the rate region described by (28) is equivalent to that one given by (14). First notice that one can define the sequence of r.v.'s $\langle X_1, X_2, \ldots, X_p \rangle$ as in (24) where $\langle U, V_1, V_2, \ldots, V_p \rangle$ is a sequence of Gaussian-distributed r.v.'s with zero mean and unit variance, independent of each other and also independent of the state matrix $S$, to satisfy the joint p.d.f. (29) and also the power constraints $\mathbb{E}_S[X_i^2] \leq P_i, i = 1, \ldots, p$. Thus by substituting $\langle U, X_1, X_2, \ldots, X_p \rangle$ as defined in (24) in the rate region described by (28) one can see that (14) results. Conversely, we show that the rate region (28) does not exceed any points of (14); first, we state some useful technical lemmas from probability theory in the following:

For arbitrary r.v.'s $X$ and $Y$, $VAR(X|Y = y)$ is defined as:

$$VAR(X|Y = y) \triangleq \mathbb{E}[X^2|Y = y] - (\mathbb{E}[X|Y = y])^2$$

(30)

in fact, it is the variance of $X$ with respect to the distribution $P_{X|Y}(x|y)$ and therefore a positive quantity.

*Lemma 2*: Consider three r.v.'s $X, Y$ and $Z$:

I) Assume that $Y$ and $Z$ are independent. The following holds:

$$\mathbb{E}_Y\big[\mathbb{E}[X^2|Y,Z=z]\big] = \mathbb{E}[X^2|Z=z]$$

(31)

II) Assume that $X$ and $Z$ are $\mathbb{R}$-valued and $X \to Y \to Z$ forms a Markov chain. Then:

$$VAR(X+Z|Y=y) = VAR(X|Y=y) + VAR(Z|Y=y).$$

(32)

*Proof:* By direct computation.

*Lemma 3:* Consider the r.v.'s $\langle S, E_1, E_2, U, X_1, X_2 \rangle$ with the joint p.d.f. given as follows:

$$P_{SE_1E_2UX_1X_2}(s,e_1,e_2,u,x_1,x_2) = P_{SE_1E_2}(s,e_1,e_2)P_U(u)\prod_{i=1}^{2} P_{X_i|UE_i}(x_i|u,e_i)$$

(33)

where $X_1$ and $X_2$ are $\mathbb{R}$-valued and there exist deterministic functions $\xi_i(.)$ such that: $E_i = \xi_i(S), i = 1,2$. Then, the following inequality holds:

$$\mathbb{E}[X_1 X_2 | S=s] \leq \sqrt{\mathbb{E}_U[(\mathbb{E}[X_1|U,E_1=\xi_1(s)])^2]}\sqrt{\mathbb{E}_U[(\mathbb{E}[X_2|U,E_2=\xi_2(s)])^2]}.$$

(34)

*Proof:* Since $S$ is independent of $U$, we can write:

$\mathbb{E}[X_1 X_2 | S=s] = \mathbb{E}_U\big[\mathbb{E}[X_1 X_2 | U, S=s]\big]$

$\stackrel{(a)}{=} \mathbb{E}_U\big[\mathbb{E}[X_1|U,S=s]\mathbb{E}[X_2|U,S=s]\big]$

$\stackrel{(b)}{=} \mathbb{E}_U\big[\mathbb{E}[X_1|U,E_1=\xi_1(s)]\mathbb{E}[X_2|U,E_2=\xi_2(s)]\big]$

$\stackrel{(c)}{\leq} \sqrt{\mathbb{E}_U[(\mathbb{E}[X_1|U,E_1=\xi_1(s)])^2]}\sqrt{\mathbb{E}_U[(\mathbb{E}[X_2|U,E_2=\xi_2(s)])^2]}$

(35)

where (a) is true because $X_2 \to (U,S) \to X_1$ forms a Markov chain, (b) is true since $S \to \big(U, E_1=\xi_1(S)\big) \to X_1$ and also $S \to \big(U, E_2=\xi_2(S)\big) \to X_2$ form Markov chains, and finally (c) is due to the Cauchy-Schwarz inequality.

Now, consider the rate region described in (28). For $i = 1, \dots, \wp$, define deterministic functions $\varphi_i(.): \mathcal{E}_i \to \mathbb{R}^+$ such that:

$$\forall e_i \in \mathcal{E}_i: \quad \varphi_i(e_i) \triangleq \mathbb{E}[X_i^2|E_i=e_i]$$

(36)

and also deterministic functions $\varrho_i(.): \mathcal{E}_i \to [0,1]$ as follows:

$$\forall e_i \in \mathcal{E}_i: \quad \varrho_i(e_i) \triangleq \sqrt{\frac{\mathbb{E}_U[(\mathbb{E}[X_i|U,E_i=e_i])^2]}{\varphi_i(e_i)}}.$$

(37)

Notice that:

$$\mathbb{E}[X_i^2|U=u,E_i=e_i] - (\mathbb{E}[X_i|U=u,E_i=e_i])^2 = VAR(X_i|U=u,E_i=e_i) \geq 0,$$

(38)

which yields:

$$\mathbb{E}[X_i^2|E_i=e_i] - \mathbb{E}_U[(\mathbb{E}[X_i|U,E_i=e_i])^2] \stackrel{(a)}{=} \mathbb{E}_U[\mathbb{E}[X_i^2|U,E_i=e_i] - (\mathbb{E}[X_i|U,E_i=e_i])^2] \geq 0$$

(39)

where (a) is obtained from Part I of Lemma 2. Therefore, for each $e_i \in \mathcal{E}_i$, $\varrho_i(e_i)$ belongs to the interval $[0,1]$ and the notation $\varrho_i(.): \mathcal{E}_i \to [0,1]$ is well-defined. Furthermore, from (36) we have:

$$\mathbb{E}[\varphi_i(E_i)] = \mathbb{E}\big[\mathbb{E}[X_i^2|E_i]\big] = \mathbb{E}[X_i^2] \leq P_i, \quad i = 1, \dots, p.$$

(40)

Now for the mutual information function terms in the rate region (28) we can proceed as:

$\forall \Lambda \subseteq \{1, \dots, p\}, \forall j \in \{1, \dots, q\}$,

$$I(\{X_k\}_{k\in\Lambda}; Y_j | \{X_k\}_{k\notin\Lambda}, U, S) = h\left(\sum_{k=1}^{p} S_{jk}X_k + Z_j \bigg| \{X_k\}_{k\notin\Lambda}, U, S\right) - h\left(\sum_{k=1}^{p} S_{jk}X_k + Z_j \bigg| \{X_k\}_{k=1}^{p}, U, S\right)$$

$$\stackrel{(a)}{=} h\left(\sum_{k\in\Lambda} S_{jk}X_k + Z_j \bigg| U, S\right) - \frac{1}{2}\log(2\pi e \sigma_j^2)$$

$$\stackrel{(b)}{=} \mathbb{E}_S\left[\mathbb{E}_U\left[h\left(\sum_{k\in\Lambda} S_{jk}X_k + Z_j \bigg| U=u, S=s\right)\right]\right] - \frac{1}{2}\log(2\pi e \sigma_j^2)$$

$$\leq \mathbb{E}_S\left[\mathbb{E}_U\left[\frac{1}{2}\log\left(2\pi e VAR\left(\sum_{k\in\Lambda} S_{jk}X_k + Z_j \bigg| U=u, S=s\right)\right)\right]\right] - \frac{1}{2}\log(2\pi e \sigma_j^2)$$

$$= \mathbb{E}_S\left[\mathbb{E}_U\left[\frac{1}{2}\log\left(2\pi e \left(\sum_{k\in\Lambda}(S_{jk})^2 VAR(X_k|U=u, S=s) + \sigma_j^2\right)\right)\right]\right] - \frac{1}{2}\log(2\pi e \sigma_j^2)$$

$$\stackrel{(c)}{=} \mathbb{E}_S\left[\mathbb{E}_U\left[\mathbb{C}\left(\frac{\sum_{k\in\Lambda}(S_{jk})^2 VAR(X_k|U=u, E_k=e_k)}{\sigma_j^2}\right)\right]\right]$$

$$= \mathbb{E}_S\left[\mathbb{E}_U\left[\mathbb{C}\left(\frac{\sum_{k\in\Lambda}(S_{jk})^2 (\mathbb{E}[X_k^2|U=u,E_k=e_k] - (\mathbb{E}[X_i|U=u,E_i=e_i])^2)}{\sigma_j^2}\right)\right]\right]$$

$$\stackrel{(d)}{\leq} \mathbb{E}_S\left[\mathbb{C}\left(\frac{\sum_{k\in\Lambda}(S_{jk})^2 \mathbb{E}_U[\mathbb{E}[X_k^2|U=u,E_k=e_k] - (\mathbb{E}[X_i|U=u,E_i=e_i])^2]}{\sigma_j^2}\right)\right]$$

$$\stackrel{(f)}{=} \mathbb{E}_S\left[\mathbb{C}\left(\frac{\sum_{k\in\Lambda}\left((S_{kj})^2 \varphi_k(E_k)\left(1-\varrho_k^2(E_k)\right)\right)}{\sigma_j^2}\right)\right]$$

(41)

where (a) is true since from the joint p.d.f. (29), $\{X_k\}_{k\in\Lambda} \to (U,S) \to \{X_k\}_{k\notin\Lambda}$ forms a Markov chain, (b) is true since $U$ and $S$ are independent of each other, (c) is due to the fact that from (29) $S \to (U, E_k) \to X_k$ forms a Markov chain, (d) is due to Jensen's inequality and (f) is obtained from definitions (36) and (37). Moreover, for the bounds on the sum rate we have:

$$I(X_1, \dots, X_p; Y_j | S) = h\left(\sum_{k=1}^{p} S_{jk}X_k + Z_j \bigg| S\right) - h\left(\sum_{k=1}^{p} S_{jk}X_k + Z_j \bigg| \{X_k\}_{k=1}^{p}, S\right)$$

$$= h\left(\sum_{k=1}^{p} S_{jk}X_k + Z_j \bigg| S\right) - \frac{1}{2}\log(2\pi e \sigma_j^2)$$

$$= \mathbb{E}_S\left[h\left(\sum_{k=1}^{p} S_{jk}X_k + Z_j \bigg| S = s\right)\right] - \frac{1}{2}\log(2\pi e \sigma_j^2)$$

$$\leq \mathbb{E}_S\left[\frac{1}{2}\log\left(2\pi e \mathbb{E}\left[\left(\sum_{k=1}^{p} S_{jk}X_k + Z_j\right)^2 \bigg| S = s\right]\right)\right] - \frac{1}{2}\log(2\pi e \sigma_j^2)$$

$$= \mathbb{E}_S\left[\mathbb{C}\left(\frac{\sum_{k=1}^{p}(S_{jk})^2 \mathbb{E}[X_k^2|S=s] + 2\sum_{1\leq k<i\leq p}(S_{jk}S_{ji}\mathbb{E}[X_k X_i|S=s])}{\sigma_j^2}\right)\right]$$

$$\stackrel{(a)}{\leq} \mathbb{E}_S\left[\mathbb{C}\left(\frac{\sum_{k=1}^{p}(S_{jk})^2 \mathbb{E}[X_k^2|E_k=\xi_k(s)] + 2\sum_{1\leq k<i\leq p}\left(S_{jk}S_{ji}\sqrt{\mathbb{E}_U[(\mathbb{E}[X_k|U,E_k=\xi_k(s)])^2]}\sqrt{\mathbb{E}_U[(\mathbb{E}[X_i|U,E_i=\xi_i(s)])^2]}\right)}{\sigma_j^2}\right)\right]$$

$$\stackrel{(b)}{=} \mathbb{E}_S\left[\mathbb{C}\left(\frac{\sum_{k=1}^{p}(S_{jk})^2 \varphi_k(E_k) + 2\sum_{1\leq k<i\leq p}\left(S_{jk}S_{ji}\varrho_k(E_k)\varrho_i(E_i)\sqrt{\varphi_k(E_k)\varphi_i(E_i)}\right)}{\sigma_j^2}\right)\right]$$

(42)

where (a) is obtained from Lemma 3 and also because by the joint p.d.f. (29) $S \to E_k \to X_k$ forms a Markov chain. The equality (b) is obtained from (36) and (37). Now by substituting (41) and (42) in (28) we obtain the fact that no point outside the rate region (14) is achievable. This completes the proof of Theorem 2. ∎

As mentioned in introduction, the capacity region of the Gaussian fading two transmitter/one receiver MAC with common message where perfect CSI is available at both transmitters and the receiver was previously characterized in [18]. It is clear that the result of [18] is a special case of Theorem 1. On the other hand, it can be verified that the techniques used in [18] to derive its result are different from ours, yielding different formulation of the capacity region. Nonetheless, it is expected that these two formulations are equivalent. Let us consider the two-transmitter/one-receiver Gaussian fading MAC with a common message formulated as (10), where for notational convenience we set $\sigma^2 \equiv 1$. Moreover, assume that both transmitters and also the receiver have access to perfect CSI. Using Theorem 1, the capacity region under these conditions is given by:

$$\bigcup_{\substack{\varphi_i(.):\mathbb{R}_+^2 \to \mathbb{R}_+ \\ \varrho_i(.):\mathbb{R}_+^2 \to [0,1] \\ i=1,2}} \left\{ \begin{array}{l} (R_0, R_1, R_2) \in \mathbb{R}_+^3 \\ R_1 \leq \mathbb{E}_S\left[\mathbb{C}\left((S_1)^2 \varphi_1(S)(1-\varrho_1^2(S))\right)\right] \\ R_2 \leq \mathbb{E}_S\left[\mathbb{C}\left((S_2)^2 \varphi_2(S)(1-\varrho_2^2(S))\right)\right] \\ R_1 + R_2 \leq \mathbb{E}_S\left[\mathbb{C}\left(\sum_{k=1}^{2}(S_k)^2 \varphi_k(S)(1-\varrho_k^2(S))\right)\right] \\ R_0 + R_1 + R_2 \leq \mathbb{E}_S\left[\mathbb{C}\left(\begin{array}{l}(S_1)^2 \varphi_1(S) + (S_2)^2 \varphi_2(S) \\ +2S_1 S_2 \varrho_1(S)\varrho_2(S)\sqrt{\varphi_1(S)\varphi_2(S)}\end{array}\right)\right] \end{array} \right\}$$

(43)

where $S = (S_1, S_2) \in \mathbb{R}_+^2$ is a r.v. representing the state process of the channel, $\varphi_i(.): \mathbb{R}_+^2 \to \mathbb{R}_+$ is the power allocation policy for the $i^{th}$ transmitter satisfying the following constraint:

$$\mathbb{E}[\varphi_i(S)] \leq P_i, \quad i = 1,2$$

(44)

and $\varrho_i(.): \mathbb{R}_+^2 \to [0,1]$ is an arbitrary deterministic function. This is a consequence of Theorem 1 by setting $p \equiv 2$, $q \equiv 1$, and $E_i \equiv S = (S_1, S_2), i = 1,2$, (perfect CSIT) in (14).

Now, let us bring out the formulation of [18] for the capacity region of the underlying channel, i.e., the two-transmitter/one-receiver Gaussian fading MAC with a common message with perfect CSIT and perfect CSIR. Using our notations, this formulation is given by:

$$\bigcup_{\substack{\varrho(.):\mathbb{R}_+^2 \to [0,1] \\ p_i(.):\mathbb{R}_+^2 \to \mathbb{R}_+ \\ i=0,1,2}} \begin{cases} (R_0, R_1, R_2) \in \mathbb{R}_+^3 \\ R_1 \leq \mathbb{E}_S[\mathbb{C}(p_1(H))] \\ R_2 \leq \mathbb{E}_S[\mathbb{C}(p_2(H))] \\ R_1 + R_2 \leq \mathbb{E}_S[\mathbb{C}(p_1(H) + p_2(H))] \\ R_0 + R_1 + R_2 \leq \mathbb{E}_S[\mathbb{C}(p_1(H) + p_2(H) + p_0(H))] \end{cases}$$

(45)

where $S = (S_1, S_2) \in \mathbb{R}_+^2$, $H = (S_1^2, S_2^2) \in \mathbb{R}_+^2$, and the functions $\varrho(.): \mathbb{R}_+^2 \to [0,1]$ and $p_i(.): \mathbb{R}_+^2 \to \mathbb{R}_+$, $i = 0,1,2$, satisfy the following conditions:

$$\begin{cases} \mathbb{E}_S\left[\dfrac{p_1(H)}{S_1^2} + \dfrac{\varrho(H)^2 p_0(H)}{S_1^2}\right] \leq P_1 \\ \mathbb{E}_S\left[\dfrac{p_2(H)}{S_2^2} + \dfrac{\big(1 - \varrho(H)\big)^2 p_0(H)}{S_2^2}\right] \leq P_2 \end{cases}$$

(46)

As we see, the formulation (43) is completely different from (45). Next, we prove that these two formulations describe the same rate region.

*Proposition 2:* The rate region described by (43) is equivalent to (45)-(46).

*Proof of Proposition 2:* First consider the rate region (43). Let $\varphi_i(.): \mathbb{R}_+^2 \to \mathbb{R}_+, i = 1,2$, be two arbitrary deterministic functions satisfying $\mathbb{E}[\varphi_i(S)] \leq P_i, i = 1,2$. Also, assume that $\varrho_i(.): \mathbb{R}_+^2 \to [0,1], i = 1,2,$ are two arbitrary deterministic functions. Define the functions $p_i(.): \mathbb{R}_+^2 \to \mathbb{R}_+, i = 0,1,2,$ and $\varrho(.): \mathbb{R}_+^2 \to [0,1]$ as follows:

$$\begin{cases} p_0\big(H = (S_1^2, S_2^2)\big) \triangleq \Big(S_1 \varrho_1(S)\sqrt{\varphi_1(S)} + S_2 \varrho_2(S)\sqrt{\varphi_2(S)}\Big)^2 \\ p_1\big(H = (S_1^2, S_2^2)\big) \triangleq (S_1)^2 \varphi_1(S)\big(1 - \varrho_1^2(S)\big) \\ p_2\big(H = (S_1^2, S_2^2)\big) \triangleq (S_2)^2 \varphi_2(S)\big(1 - \varrho_2^2(S)\big) \\ \varrho\big(H = (S_1^2, S_2^2)\big) \triangleq \dfrac{S_1 \varrho_1(S)\sqrt{\varphi_1(S)}}{S_1 \varrho_1(S)\sqrt{\varphi_1(S)} + S_2 \varrho_2(S)\sqrt{\varphi_2(S)}} \end{cases}$$

(47)

Note that $S = (S_1, S_2) = \left(\sqrt{S_1^2}, \sqrt{S_2^2}\right)$. It can be easily verified that the functions $p_i(.), i = 0,1,2,$ and $\varrho(.)$ defined by (47) satisfy (46). Now by substituting $p_i(.), i = 0,1,2,$ and $\varrho(.)$ as defined by (47), in (45) we obtain that the rate region (43) is a subset of (45). Conversely, consider the rate region (45). Let $p_i(.): \mathbb{R}_+^2 \to \mathbb{R}_+, i = 0,1,2,$ and $\varrho(.): \mathbb{R}_+^2 \to [0,1]$ be arbitrary deterministic functions that satisfy (46). Define the functions $\varphi_i(.): \mathbb{R}_+^2 \to \mathbb{R}_+, i = 1,2,$ and $\varrho_i(.): \mathbb{R}_+^2 \to [0,1], i = 1,2,$ as follows:

$$\begin{cases} \varphi_1\big(S = (S_1, S_2)\big) \triangleq \dfrac{p_1(S_1^2, S_2^2)}{S_1^2} + \dfrac{\varrho(S_1^2, S_2^2)^2 p_0(S_1^2, S_2^2)}{S_1^2} \\ \varphi_2\big(S = (S_1, S_2)\big) \triangleq \dfrac{p_2(S_1^2, S_2^2)}{S_2^2} + \dfrac{\big(1 - \varrho(S_1^2, S_2^2)\big)^2 p_0(S_1^2, S_2^2)}{S_2^2} \\ \varrho_1\big(S = (S_1, S_2)\big) \triangleq \dfrac{\varrho(S_1^2, S_2^2)\sqrt{p_0(S_1^2, S_2^2)}}{\sqrt{p_1(S_1^2, S_2^2) + \varrho(S_1^2, S_2^2)^2 p_0(S_1^2, S_2^2)}} \\ \varrho_2\big(S = (S_1, S_2)\big) \triangleq \dfrac{\big(1 - \varrho(S_1^2, S_2^2)\big)\sqrt{p_0(S_1^2, S_2^2)}}{\sqrt{p_2(S_1^2, S_2^2) + \big(1 - \varrho(S_1^2, S_2^2)\big)^2 p_0(S_1^2, S_2^2)}} \end{cases}$$

(48)

From (46), it is easily seen that $\mathbb{E}[\varphi_i(S)] \leq P_i, i = 1,2$. Thus, we can substitute $\varphi_i(.): \mathbb{R}_+^2 \to \mathbb{R}_+, i = 1,2$, and $\varrho_i(.): \mathbb{R}_+^2 \to [0,1], i = 1,2$, as defined by (48), in (43) and then find (45), verifying that (45) is a subset of (43). This completes the proof. ∎

In fact, Theorem 1 (specialized for the case of a two-transmitter/one-receiver MAC with perfect CSI at both transmitters and the receiver) together with Proposition 2 result in an alternative proof of the capacity region for the Gaussian fading MAC with a common message studied in [18].

*III-B) Two-transmitter one-receiver multiple access channel with conferencing encoders*

In this section, we consider the MAC with conferencing encoders. The capacity region of a discrete two-user MAC with conferencing encoders was obtained in [7]. In the sequence, we first state the result of Willems [7] and then derive the capacity region of a two-user Gaussian fading MAC with conferencing encoders.

*Proposition 3* [7]: Consider a discrete two-user memoryless MAC with encoders connected by communication links of capacities $C_{12}$ and $C_{21}$, as depicted in Fig. 2. The capacity region denoted by $\mathcal{C}_{conf}^{D-MAC}(C_{12}, C_{21})$, is given by:

$$\mathcal{C}_{conf}^{D-MAC}(C_{12}, C_{21}) = \bigcup_{P_U P_{X_1|U} P_{X_2|U}} \begin{Bmatrix} (R_1, R_2) \in \mathbb{R}_+^2 \\ R_1 \leq I(X_1; Y|X_2, U) + C_{12} \\ R_2 \leq I(X_2; Y|X_1, U) + C_{21} \\ R_1 + R_2 \leq I(X_1, X_2; Y|U) + C_{12} + C_{21} \\ R_1 + R_2 \leq I(X_1, X_2; Y) \end{Bmatrix}$$

(49)

To prove the achievability of the rate region in (49) for the two-user MAC with conferencing encoders, Willems [7] used a two-user discrete MAC with a common message. The capacity region of such a channel had already been obtained in [4] by Slepian and Wolf, (see also Remark 3 of Theorem 1). In the following theorem, we exploit the technique proposed by Willems [7] to obtain the capacity region of the two-user Gaussian fading MAC with conferencing encoders.

*Theorem 2:* The capacity region of the two-user Gaussian fading MAC (10) with conferencing encoders connected to each other with links of capacities $C_{12}$ and $C_{21}$, denoted by $\mathcal{C}_{conf}^{Gf-MAC}(C_{12}, C_{21})$, when the transmitters have access to partial CSIT and the receiver has access to perfect CSIR, is given by:

$$\mathcal{C}_{conf}^{Gf-MAC}(C_{12}, C_{21}) = \bigcup_{\substack{\varphi_1(.),\varphi_2(.) \\ \varrho_1(.),\varrho_2(.)}} \begin{cases} (R_1, R_2) \in \mathbb{R}_+^2 \\ R_1 \leq \mathbb{E}_S\left[\mathbb{C}\left(\frac{(S_1)^2 \varphi_1(E_1)\left(1-\varrho_1^2(E_1)\right)}{\sigma^2}\right)\right] + C_{12} \\ R_2 \leq \mathbb{E}_S\left[\mathbb{C}\left(\frac{(S_2)^2 \varphi_2(E_2)\left(1-\varrho_2^2(E_2)\right)}{\sigma^2}\right)\right] + C_{21} \\ R_1 + R_2 \leq \mathbb{E}_S\left[\mathbb{C}\left(\frac{\sum_{k=1}^2 (S_k)^2 \varphi_k(E_k)\left(1-\varrho_k^2(E_k)\right)}{\sigma^2}\right)\right] + C_{12} + C_{21} \\ R_1 + R_2 \leq \mathbb{E}_S\left[\mathbb{C}\left(\frac{(S_1)^2 \varphi_1(E_1) + (S_2)^2 \varphi_2(E_2)}{\sigma^2} + \frac{2 S_1 S_2 \varrho_1(E_1) \varrho_2(E_2) \sqrt{\varphi_1(E_1) \varphi_2(E_2)}}{\sigma^2}\right)\right] \end{cases}$$

(50)

where $S = (S_1, S_2) \in (\mathbb{R}^+)^2$ is a r.v. representing the state process of the channel and $E_i \in \mathcal{E}_i$, is a r.v. representing CSIT available at the $i^{th}$ transmitter, $i = 1,2$, which is a deterministic function of $S$, i.e. $E_i = \xi_i(S)$. Furthermore, $\varphi_i(.): \mathcal{E}_i \to \mathbb{R}^+$ is the power allocation policy for the $i^{th}$ transmitter, that satisfies the following constraint:

$$\mathbb{E}[\varphi_i(E_i)] \leq P_i, \qquad i = 1,2$$

(51)

and $\varrho_i(.): \mathcal{E}_i \to [0,1]$ is an arbitrary (limited) deterministic function that takes its values from the interval $[0,1]$.

*Proof of Theorem 2:* To prove the achievability of (50), we directly apply Willems' approach [7] for the discrete model to the Gaussian fading channel. Consider a block length-$n$ code with $2^{nR_1}$ messages $w_1$ for transmitter 1 and $2^{nR_2}$ messages $w_2$ for transmitter 2. Partition the set $\{1, ..., 2^{nR_1}\}$ into $2^{nR_{12}}$ cells each containing $2^{n(R_1 - R_{12})}$ elements. We label the cells $c_1 \in \{1, ..., 2^{nR_{12}}\}$ and the elements inside each cell $t_1 \in \{1, ..., 2^{n(R_1 - R_{12})}\}$. Let $c_1(w_1) = s_1$, if $w_1$ is inside the cell $s_1$. A similar partitioning is done for the message set $\{1, ..., 2^{nR_2}\}$. We define $R_{12} = \min\{R_1, C_{12}\}$ and $R_{21} = \min\{R_2, C_{21}\}$. Now since $R_{12} \leq C_{12}$ and $R_{21} \leq C_{21}$, it is possible for encoder 1 to send $c_1(w_1) \in \{1, ..., 2^{nR_{12}}\}$ to encoder 2 and for encoder 2 to send $c_2(w_2) \in \{1, ..., 2^{nR_{21}}\}$ to encoder 1, by holding a $(C_{12}, C_{21})$-permissible conference. Thus, one can define a common message for the encoders as:

$$\acute{w}_0 = (c_1(w_1), c_2(w_2)).$$

We also note that $t_1$ and $t_2$ are unknown to encoders 2 and 1, respectively. Therefore, $\acute{w}_0 \in \{1, ..., 2^{n(R_{12} + R_{21})}\}$ can be considered as a common message and $t_1 \in \{1, ..., 2^{n(R_1 - R_{12})}\}$ and $t_2 \in \{1, ..., 2^{n(R_2 - R_{21})}\}$ as private messages. Now, from the result of Theorem 1 specialized for the case of a two-user Gaussian fading MAC with a common message, we conclude that $(R_1, R_2)$ is achievable for the two-user Gaussian fading MAC with conferencing encoders if the rates $R_1, R_2, R_{12}, R_{21}$, satisfy the following:

$$\begin{cases} 0 \leq R_1 - R_{12} \leq \mathbb{E}_S\left[\mathbb{C}\left(\frac{(S_1)^2 \varphi_1(E_1)\left(1-\varrho_1^2(E_1)\right)}{\sigma^2}\right)\right] \\ 0 \leq R_2 - R_{21} \leq \mathbb{E}_S\left[\mathbb{C}\left(\frac{(S_2)^2 \varphi_2(E_2)\left(1-\varrho_2^2(E_2)\right)}{\sigma^2}\right)\right] \\ R_1 + R_2 - (R_{12} + R_{21}) \leq \mathbb{E}_S\left[\mathbb{C}\left(\frac{\sum_{k=1}^2 (S_k)^2 \varphi_k(E_k)\left(1-\varrho_k^2(E_k)\right)}{\sigma^2}\right)\right] \\ R_1 + R_2 \leq \mathbb{E}_S\left[\mathbb{C}\left(\frac{(S_1)^2 \varphi_1(E_1) + (S_2)^2 \varphi_2(E_2) + 2 S_1 S_2 \varrho_1(E_1) \varrho_2(E_2) \sqrt{\varphi_1(E_1) \varphi_2(E_2)}}{\sigma^2}\right)\right] \end{cases}$$

(52)

for some power allocation policy functions $\varphi_i(.), i = 1,2$, satisfying (51) and also two arbitrary deterministic functions $\varrho_i: \mathcal{E}_i \to [0,1], i = 1,2$. On the other hand, by definition of $R_{12}$ and $R_{21}$, one can easily see that the rates $R_1, R_2, R_{12}, R_{21}$, satisfy (52), if and only if $(R_1, R_2) \in \mathcal{C}_{conf}^{Gf-MAC}(C_{12}, C_{21})$, where $\mathcal{C}_{conf}^{Gf-MAC}(C_{12}, C_{21})$ is given by (50). This proves the achievability of (50). For the converse part similar to Theorem 1, we first derive an outer bound on $\mathcal{C}_{conf}^{Gf-MAC}(C_{12}, C_{21})$ in terms of mutual information functions, in the following lemma.

*Lemma 4:* The capacity region of the two-user Gaussian fading MAC with conferencing encoders, i.e., $\mathcal{C}_{conf}^{Gf-MAC}(C_{12}, C_{21})$, is outer bounded by:

$$\mathcal{C}_{conf}^{Gf-MAC}(C_{12}, C_{21}) \subseteq \bigcup_{\substack{P_U P_{X_1|UE_1} P_{X_2|UE_2} \\ \mathbb{E}[X_i^2] \leq P_i, i=1,2}} \begin{cases} (R_1, R_2) \in \mathbb{R}_+^2 \\ R_1 \leq I(X_1; Y|X_2, U, S) + C_{12} \\ R_2 \leq I(X_2; Y|X_1, U, S) + C_{12} \\ R_1 + R_2 \leq I(X_1, X_2; Y|U, S) + C_{12} + C_{21} \\ R_1 + R_2 \leq I(X_1, X_2; Y|S) \end{cases}$$

(53)

where $S = (S_1, S_2) \in (\mathbb{R}^+)^2$ is a r.v. representing the state process of the channel and $E_i \in \mathcal{E}_i$, is a r.v. representing CSIT available at the $i^{th}$ transmitter, which is a deterministic function of $S$: $E_i = \xi_i(S), i = 1,2$.

*Proof:* See Appendix III.

Now, note that the mutual information functions in the rate region described by (53) and also the joint p.d.f. over which the union in (53) is taken are exactly the same as (28), (29), respectively, when specialized for the case of a two-user MAC. Thus, one can proceed as in proof of Theorem 1 to optimize the rate region given in (53) using suitable joint p.d.f.'s, and show that the rate region given in (53) is equivalent to $\mathcal{C}_{conf}^{Gf-MAC}(C_{12}, C_{21})$ given by (50). This completes the proof of Theorem 2. ∎

## IV. NUMERICAL EXAMPLES AND SIMULATIONS

In this section, we provide some numerical results for the Gaussian fading two transmitter/one receiver MACs studied in the previous section. We consider the Gaussian fading channel with no CSIT (the transmitters have no knowledge of CSI) and perfect CSIR. At first, we briefly review the results obtained in previous section under these assumptions, in the following.

*Corollary:* Consider a two-user Gaussian fading MAC (10) with no CSIT and perfect CSIR:

I) The capacity region of the channel with a common message, denoted by $\mathcal{C}_{cm}^{no-CSIT}$, is given by:

$$\mathcal{C}_{cm}^{no-CSIT} = \bigcup_{0 \leq \varrho_1, \varrho_2 \leq 1} \begin{cases} (R_0, R_1, R_2) \in \mathbb{R}_+^3 \\ R_1 \leq \mathbb{E}_S\left[\mathbb{C}\left(\frac{(S_1)^2 P_1(1-\varrho_1^2)}{\sigma^2}\right)\right] \\ R_2 \leq \mathbb{E}_S\left[\mathbb{C}\left(\frac{(S_2)^2 P_2(1-\varrho_2^2)}{\sigma^2}\right)\right] \\ R_1 + R_2 \leq \mathbb{E}_S\left[\mathbb{C}\left(\frac{(S_1)^2 P_1(1-\varrho_1^2)+(S_2)^2 P_2(1-\varrho_2^2)}{\sigma^2}\right)\right] \\ R_0 + R_1 + R_2 \leq \mathbb{E}_S\left[\mathbb{C}\left(\frac{(S_1)^2 P_1+(S_2)^2 P_2+2S_1 S_2 \varrho_1 \varrho_2 \sqrt{P_1 P_2}}{\sigma^2}\right)\right] \end{cases}$$

(54)

II) The capacity region of the channel model with connected encoders by links of capacities $C_{12}$ and $C_{21}$, denoted by $\mathcal{C}_{conf}^{no-CSIT}(C_{12}, C_{21})$, is given by:

$$\mathcal{C}_{conf}^{no-CSIT}(C_{12}, C_{21}) = \bigcup_{0 \leq \varrho_1, \varrho_2 \leq 1} \begin{cases} (R_0, R_1, R_2) \in \mathbb{R}_+^3 \\ R_1 \leq \mathbb{E}_S\left[\mathbb{C}\left(\frac{(S_1)^2 P_1(1-\varrho_1^2)}{\sigma^2}\right)\right] + C_{12} \\ R_2 \leq \mathbb{E}_S\left[\mathbb{C}\left(\frac{(S_2)^2 P_2(1-\varrho_2^2)}{\sigma^2}\right)\right] + C_{21} \\ R_1 + R_2 \leq \mathbb{E}_S\left[\mathbb{C}\left(\frac{(S_1)^2 P_1(1-\varrho_1^2)+(S_2)^2 P_2(1-\varrho_2^2)}{\sigma^2}\right)\right] + C_{12} + C_{21} \\ R_1 + R_2 \leq \mathbb{E}_S\left[\mathbb{C}\left(\frac{(S_1)^2 P_1+(S_2)^2 P_2+2S_1 S_2 \varrho_1 \varrho_2 \sqrt{P_1 P_2}}{\sigma^2}\right)\right] \end{cases}$$

(55)

*Proof:* Part I is obtained from Theorem 1 by assuming a two-user MAC and also setting $E_k \equiv \emptyset, k = 1,2$. For part II set $E_k \equiv \emptyset, k = 1,2$, in $\mathcal{C}_{conf}^{Gf-MAC}(C_{12}, C_{21})$ given by (50). ∎

Now, we examine a few implications of our results for the channel model in (10), for a Rayleigh fading environment. Assume that the fading processes $\{S_{1,t}\}_{t=1}^{\infty}$ and $\{S_{2,t}\}_{t=1}^{\infty}$ in (10) are independent of each other and $S_{k,t}, k = 1,2$, is a Rayleigh-distributed r.v. with a p.d.f. given by:

$$P_{S_k}(s) = 2se^{-s^2}, \qquad s \geq 0 \tag{56}$$

In the following, our discussion is concerned to the systems with capacities given by (54) and (55), where the state process of the channel is described by (56).

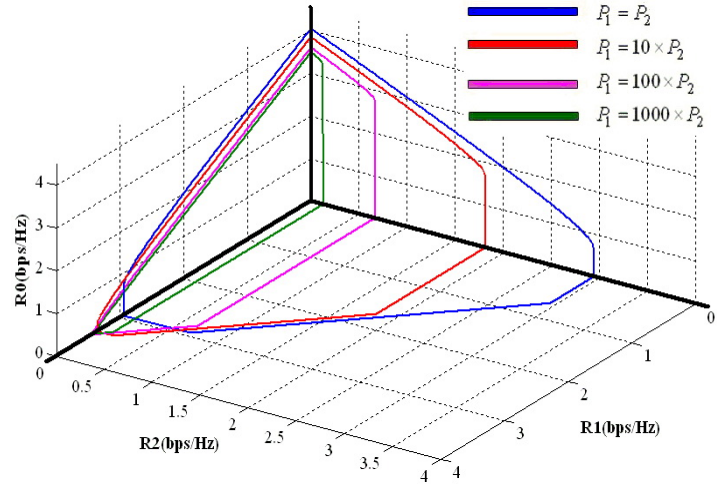

Figure 3. The capacity region of the two-user Gaussian fading MAC with common message, $\mathcal{C}_{cm}^{no-CSIT}$. The sum power for all plots is fixed and is equal to $P_1 + P_2 = 23.01$ dB, and also $P_z = 0$ dB.

We have plotted the capacity region of the two-user Gaussian MAC with common message, $\mathcal{C}_{cm}^{no-CSIT}$, in Fig. 3, under different values of the power ratio of transmitters $\frac{P_1}{P_2}$ (boundaries in each plane with solid lines). The sum power of the two transmitters, i.e., $P_1 + P_2$ is fixed for all plots. As we see from Fig. 3, when both transmitters have the same power, i.e., $\frac{P_1}{P_2} = 1$, the maximum rate of $R_0$ is attained. This can be justified as follows. It could be easily verified from (54) that the maximum achievable rate of the common message $R_0$ is given by:

$$\max_{R_0 \in \mathcal{C}_{cm}^{no-CSIT}} R_0 = \mathbb{E}_S\left[\mathbb{C}\left(\frac{(S_1)^2 P_1 + (S_2)^2 P_2 + 2S_1 S_2 \sqrt{P_1 P_2}}{\sigma^2}\right)\right] \tag{57}$$

On the other hand, for fixed value of $P_1 + P_2$, the geometric average $\sqrt{P_1 P_2}$ is maximum when $P_1 = P_2$, (note that $S_1$ and $S_2$ are positive-valued i.i.d. r.v.'s). Therefore, when the powers $P_1$ and $P_2$ become far from each other, the maximum achievable value of the rate $R_0$, i.e., the cut-off point on the $R_0$ axis, decreases. Note that for values of $\frac{P_1}{P_2}$ that are greater than 10, the variation of the power $P_1$ is less than 10 percent, therefore the increment in the maximum achievable rate of $R_1$, i.e., the cut-off point on the $R_1$-axis, is not significant, as can be seen from Fig 3.

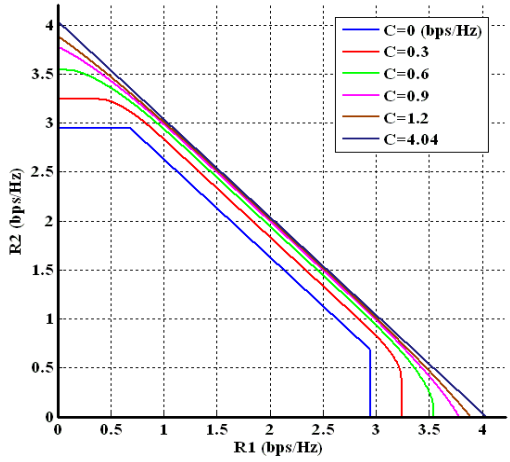

Figure 4. The capacity region of the two-user Gaussian fading MAC with conferencing encoders, $C_{conf}^{no-CSIT}(C_{12}, C_{21})$, for $C_{12} = C_{21} \triangleq C$, and $P_1 = P_2 \triangleq P = 20$ dB.

In Fig. 4, the capacity region of the two-user Gaussian fading MAC with conferencing encoders, i.e., $C_{conf}^{no-CSIT}(C_{12}, C_{21})$ is plotted for the case $C_{12} = C_{21} = C$, and $P_1 = P_2 = P$ (boundaries with solid lines). When there is no conferencing, i.e. $C = 0$, the capacity region forms a pentagon. It is clear from the figure that as $C$ increases beyond zero the capacity region enlarges. However, no further improvement is possible for values of $C$ such that $C \geq \mathbb{E}_S\left[\mathbb{C}\left(\frac{(S_1)^2 P_1 + (S_2)^2 P_2 + 2 S_1 S_2 \sqrt{P_1 P_2}}{\sigma^2}\right)\right]$, (in Fig. 4, for $C \geq 4.04$ (bps/Hz)). In fact, for values of $C$ equal or greater than $\mathbb{E}_S\left[\mathbb{C}\left(\frac{(S_1)^2 P_1 + (S_2)^2 P_2 + 2 S_1 S_2 \sqrt{P_1 P_2}}{\sigma^2}\right)\right]$, the capacity region forms a triangle whose boundary lines are given by $R_1 = 0$, $R_2 = 0$, and also:

$$R_1 + R_2 = \mathbb{E}_S\left[\mathbb{C}\left(\frac{(S_1)^2 P_1 + (S_2)^2 P_2 + 2 S_1 S_2 \sqrt{P_1 P_2}}{\sigma^2}\right)\right]$$

(58)

In this case, the conferencing rate $C$ is high enough such that each encoder can perfectly communicate the intended message to the other encoder by holding a conference before transmission, hence the channel reduces to a MAC in which both of the transmitters would send a same message $W = (W_1, W_2)$. Therefore, the capacity region is given by (58) which is the same as the maximum achievable rate of $R_0$ for the Gaussian fading MAC with common message given by (57), and no further improvement can be obtained by increasing the rate of conferencing.

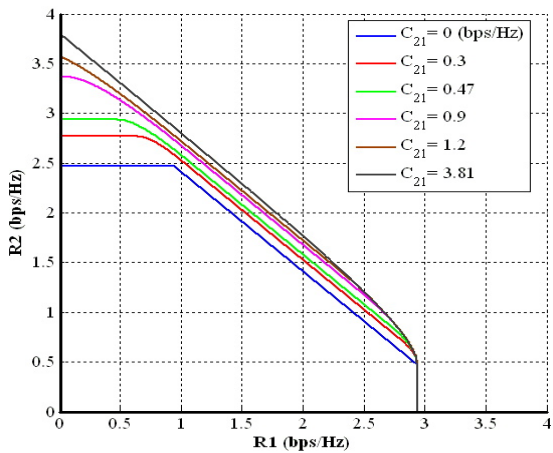

Figure 5. The capacity region of the two-user Gaussian fading MAC with one encoder communicating to the other, $C_{conf}^{no-CSIT}(0, C_{21})$, for $P_1 = 23.01$ dB, $P_2 = 20$ dB.

Now consider the case where only one of the encoders is connected to the other via a link of given capacity. In Fig. 5, the capacity region of the two-user Gaussian fading MAC with one encoder communicating to the other, i.e., $C_{conf}^{no-CSIT}(0, C_{21})$, has been plotted. Therefore, in this situation the second encoder can communicate to the first encoder via a link of capacity $C_{21}$. It is clear from the figure that as the rate of conferencing $C_{21}$ increases, the capacity region enlarges, however, since only the second encoder can cooperate its message with the first one via conferencing, the maximum achievable rate of $R_1$ (the cut-off point at $R_1$-axis), does not change. In Fig. 5 it has been assumed that the associated power of the second encoder is half of that one for the first encoder, i.e. $P_2 = \frac{1}{2} P_1$. This causes the capacity region with no conferencing ($C_{21} = 0$) to be an asymmetric region (the region with blue boundary in Fig. 5), such that the value of the cut-off point on the $R_2$-axis is less than that one on the $R_1$-axis. As we see from Fig. 5, by holding a conference with the rate of $C_{21} = 0.47$ (bps/Hz), (the region with green boundary), however, the second encoder can cooperate some part of its message to the first one, thus the maximum achievable rate of $R_2$ is the same as that one for $R_1$, i.e., the cut-off points on both axes are the same. This means that conferencing can compensate for the lack of power. In general, one can see that when the second transmitter is restricted by $P_2 = \alpha P_1$ where $0 < \alpha < 1$, a conferencing link of capacity $C_{21}(\alpha) = \mathbb{E}_S \left[ \mathbb{C} \left( \frac{(S_1)^2 \bar{\alpha} P_1}{\sigma^2 + (S_2)^2 \alpha P_2} \right) \right]$ could compensate for the lack of power, depicted by $C_{21}(0.5) = 0.47$ (bps/Hz) in Fig. 5. Moreover, as in previous setup, for values of $C_{21}$ greater than $\mathbb{E}_S \left[ \mathbb{C} \left( \frac{(S_1)^2 P_1 + (S_2)^2 P_2 + 2 S_1 S_2 \sqrt{P_1 P_2}}{\sigma^2} \right) \right]$, (in Fig. 5 for $C_{21} \geq 3.81$ (bps/Hz)), no further improvement of the capacity region is possible. In fact, for values of $C_{21}$ equal or greater than $\mathbb{E}_S \left[ \mathbb{C} \left( \frac{(S_1)^2 P_1 + (S_2)^2 P_2 + 2 S_1 S_2 \sqrt{P_1 P_2}}{\sigma^2} \right) \right]$, by holding a conference, the second transmitter can cooperate perfectly with the first one, and the channel acts as a MAC with degraded message sets: a common message is sent by both transmitters and a private message by the first transmitter. Therefore, the shape of the capacity region is similar to the capacity region of MAC with common message in $(R_0, R_1)$-plane, depicted in Fig .3.

## V. CONCLUSION

In this paper, we characterized the capacity region for some multiple-access scenarios with cooperative transmitters. Firstly, the general MAC with common information was considered. The capacity region of the discrete memoryless channel was characterized. Then, the general Gaussian fading MAC with common information with partial CSI at the transmitters and perfect CSI at the receivers was investigated. A coding theorem was proved for this model yielding an exact characterization of the throughput capacity region. Finally, the capacity region of a two-transmitter/one-receiver Gaussian fading MAC with conferencing encoders with partial CSIT and perfect CSIR was determined. For the Gaussian fading systems in a Rayleigh fading environment, some numerical examples and simulations were provided and the effect of conferencing on the improvement of the capacity region was explored. In future work, we will study the problem of optimal resource allocation for the Gaussian fading channels with partial CSIT and perfect CSIR that were considered in this paper.

## ACKNOWLEDGMENT

We would like to thank R. Bayat, M. Moghimi, the associate editor, and the anonymous reviewers for their helpful comments.

## APPENDIX I

### PROOF OF THEOREM 1

To prove achievability of (12) a length-$n$ random codebook $\mathfrak{C}_n$ is constructed as follows:

1) Fix the distribution $P_{UX_1...X_p}(u, x_1, ..., x_p) = P_U(u) \prod_{i=1}^{p} P_{X_i|U}(x_i|u)$.
2) Generate randomly $2^{nR_0}$ i.i.d. sequences $u^n$ according to $P(u^n) = \prod_{t=1}^{n} P_U(u_t)$. Label these sequences $u^n(w_0)$, where $w_0 \in \{1, ..., 2^{nR_0}\}$.
3) For each $u^n(w_0)$, generate $2^{nR_i}$, $i = \{1, 2, ..., p\}$, sequences $x_i^n$ according to the p.d.f. $P(x_i^n | u^n(w_0)) = \prod_{t=1}^{n} P_{X_i|U}(x_{i,t}|u_t)$. Label these sequences $x_i^n(w_0, w_i)$, where $w_i \in \{1, ..., 2^{nR_i}\}$.

*Encoding:* To send a common message $w_0$ and a private message $w_i$, the $i^{th}$ encoder sends the codeword $x_i^n(w_0, w_i)$ over the channel.

*Decoding* at the $j^{th}$ receiver, $j \in \{1,2,\ldots,q\}$: After receiving $y_j^n$, the $j^{th}$ receiver tries to find a unique $(\hat{w}_0, \hat{w}_1, \ldots, \hat{w}_p)$ such that $(u^n(\hat{w}_0), x_1^n(\hat{w}_0, \hat{w}_1), \ldots, x_p^n(\hat{w}_0, \hat{w}_p), y_j^n) \in A_\epsilon^n(P_{UX_1\ldots X_p Y_j})$.

*Analysis of probability of error*: From (3) we have: $P_e^{\mathfrak{C}_n} \leq P_{e,1}^{\mathfrak{C}_n} + \cdots + P_{e,q}^{\mathfrak{C}_n}$, where $P_{e,j}^{\mathfrak{C}_n}$ is given by (2). Now we compute each $P_{e,j}^{\mathfrak{C}_n}$ (error probability at the $j^{th}$ receiver). Define the events:

$$E_{i_0,i_1,\ldots,i_p}^{(j)} \triangleq \left\{(U^n(i_0), X_1^n(i_0, i_1), \ldots, X_p^n(i_0, i_p), Y_j^n) \in A_\epsilon^n(P_{UX_1\ldots X_p Y_j})\right\}, \qquad j = 1, \ldots, q.$$

By symmetry of the random code construction, the conditional probability of error does not depend on the choice of indices. Thus, the conditional probability of error is the same as the unconditional probability of error, such that without loss of generality we can assume that $(\bar{w}_0, \bar{w}_1, \ldots, \bar{w}_p)$ is sent. Then we have:

$$P_{e,j}^{\mathfrak{C}_n} = Pr\left\{\left(E_{\bar{w}_0,\bar{w}_1,\ldots,\bar{w}_p}^{(j)}\right)^c \cup \left\{\bigcup_{\substack{(i_0,i_1,\ldots,i_p) \neq \\ (\bar{w}_0,\bar{w}_1,\ldots,\bar{w}_p)}} E_{i_0,i_1,\ldots,i_p}^{(j)}\right\}\right\}$$

$$\leq Pr\left\{\left(E_{\bar{w}_0,\bar{w}_1,\ldots,\bar{w}_p}^{(j)}\right)^c\right\} + Pr\left\{\bigcup_{\substack{(i_0,i_1,\ldots,i_p) \\ i_0 \neq \bar{w}_0}} E_{i_0,i_1,\ldots,i_p}^{(j)}\right\} + \sum_{\Lambda \subseteq \{1,\ldots,p\}} \left(Pr\left\{\bigcup_{\{(i_1,\ldots,i_p) \mid \substack{i_k = \bar{w}_k \text{ if } k \notin \Lambda, \\ i_k \neq \bar{w}_k \text{ if } k \in \Lambda}\}} E_{\bar{w}_0,i_1,\ldots,i_p}^{(j)}\right\}\right)$$

(59)

from the Asymptotic Equipartition Property (AEP) theorem [19, Ch. 3], $Pr\left\{\left(E_{\bar{w}_0,\bar{w}_1,\ldots,\bar{w}_p}^{(j)}\right)^c\right\} \to 0$ as $n \to \infty$ and $\epsilon \to 0$. Furthermore, from [19, Th. 15.2.1 and 15.2.3] one can see:

$$Pr\left\{\bigcup_{\substack{(i_0,i_1,\ldots,i_p) \\ i_0 \neq \bar{w}_0}} E_{i_0,i_1,\ldots,i_p}^{(j)}\right\} \to 0,$$

(60)

Provided that $0 \leq \sum_{k=0}^{p} R_k \leq I(U, X_1, \ldots, X_p; Y_j)$, $n \to \infty$ and $\epsilon \to 0$. From this and by considering that $I(U, X_1, \ldots, X_p; Y_j) = I(X_1, \ldots, X_p; Y_j)$ due to the Markov structure $U \to X_1, \ldots, X_p \to Y_j$, the last bound in (12) is obtained. Again, from [19, Th. 15.2.1 and 15.2.3] we can see :

$$Pr\left\{\bigcup_{\{(i_1,\ldots,i_p) \mid \substack{i_k = \bar{w}_k \text{ if } k \notin \Lambda, \\ i_k \neq \bar{w}_k \text{ if } k \in \Lambda}\}} E_{\bar{w}_0,i_1,\ldots,i_p}^{(j)}\right\} \to 0,$$

(61)

Provided that $0 \leq \sum_{k \in \Lambda} R_k \leq I(\{X_k\}_{k \in \Lambda}; Y_j | \{X_k\}_{k \notin \Lambda}, U)$, $n \to \infty$ and $\epsilon \to 0$. The above bounds show that the average probability of error is arbitrarily small provided that $(R_0, R_1, R_2, \ldots, R_p)$ belongs to the rate region (12), as desired. ∎

# APPENDIX II

## PROOF OF LEMMA 1

Consider a length-$n$ code with average error probability $P_e^{(n)}$, where $P_e^{(n)} \to 0$ as $n \to \infty$. Let $\Lambda$ be a subset of $\{1, \dots, \wp\}$. Define new r.v.'s $\widetilde{U}_t, t = 1, \dots, n$, as: $\widetilde{U}_t \triangleq W_0$.

By Fano's inequality, for $j = 1, \dots, q$, we have:

$$n \sum_{k \in \Lambda} R_k = H(\{W_k\}_{k \in \Lambda}) = I(\{W_k\}_{k \in \Lambda}; Y_j^n, S^n) + H(\{W_k\}_{k \in \Lambda} | Y_j^n, S^n)$$

$$\leq I(\{W_k\}_{k \in \Lambda}; Y_j^n, S^n) + n\epsilon_n^{(j,\Lambda)} \leq I(\{W_k\}_{k \in \Lambda}; Y_j^n, S^n, W_0, \{W_k\}_{k \notin \Lambda}) + n\epsilon_n^{(j,\Lambda)}$$

$$= I(\{W_k\}_{k \in \Lambda}; Y_j^n | S^n, W_0, \{W_k\}_{k \notin \Lambda}) + n\epsilon_n^{(j,\Lambda)}$$

$$= \sum_{t=1}^n I(\{W_k\}_{k \in \Lambda}; Y_{j,t} | S^n, W_0, \{W_k\}_{k \notin \Lambda}, Y_j^{t-1}) + n\epsilon_n^{(j,\Lambda)}$$

$$= \sum_{t=1}^n \left[ h(Y_{j,t} | S^n, W_0, \{W_k\}_{k \notin \Lambda}, Y_j^{t-1}) - h(Y_{j,t} | S^n, W_0, \{W_k\}_{k \notin \Lambda}, \{W_k\}_{k \in L}, Y_j^{t-1}) \right] + n\epsilon_n^{(j,\Lambda)}$$

$$\stackrel{(a)}{=} \sum_{t=1}^n \left[ h\left(Y_{j,t} | S^n, W_0, \{W_k\}_{k \notin \Lambda}, Y_j^{t-1}, \{X_{k,t}\}_{k \notin \Lambda}\right) - h\left(Y_{j,t} | S^n, W_0, \{W_k\}_{k \notin \Lambda}, \{W_k\}_{k \in \Lambda}, Y_j^{t-1}, \{X_{k,t}\}_{k \notin \Lambda}, \{X_{k,t}\}_{k \in \Lambda}\right) \right] + n\epsilon_n^{(j,\Lambda)}$$

$$\stackrel{(b)}{\leq} \sum_{t=1}^n \left[ h\left(Y_{j,t} | \widetilde{U}_t, \{X_{k,t}\}_{k \notin \Lambda}, S_t\right) - h\left(Y_{j,t} | \widetilde{U}_t, \{X_{k,t}\}_{k \notin \Lambda}, \{X_{k,t}\}_{k \in \Lambda}, S_t\right) \right] + n\epsilon_n^{(j,\Lambda)}$$

$$= \sum_{t=1}^n I\left(\{X_{k,t}\}_{k \in \Lambda}; Y_{j,t} | \widetilde{U}_t, \{X_{k,t}\}_{k \notin \Lambda}, S_t\right) + n\epsilon_n^{(j,\Lambda)}$$

(62)

where $\epsilon_n^{(j,\Lambda)} \to 0$ as $n \to \infty$. The equality in (a) is due to the fact that $X_{k,t}$ is a deterministic function of $(W_0, W_k, E_k = \xi_k(S_t))$, the inequality in (b) holds because conditioning does not increase entropy and also $Y_{j,t}$ conditioned on $\left(\{X_{k,t}\}_{k=1}^{\wp}, S_t\right)$ is independent of $(S^{t-1}, S_t^n, W_0, \{W_k\}_{k \notin \Lambda}, \{W_k\}_{k \in \Lambda}, Y_j^{t-1})$. For the sum rate (last bound in (28)), again by Fano's inequality we have:

$$n \sum_{k=0}^{\wp} R_k = H(W_0, W_1, \dots, W_\wp) = I(W_0, W_1, \dots, W_\wp; Y_j^n, S^n) + H(W_0, W_1, \dots, W_\wp | Y_j^n, S^n)$$

$$\leq I(W_0, W_1, \dots, W_\wp; Y_j^n, S^n) + n\epsilon_n^{(j)}$$

$$= \sum_{t=1}^n I(W_0, W_1, \dots, W_\wp; Y_{j,t} | Y_j^{t-1}, S^n) + n\epsilon_n^{(j)}$$

$$= \sum_{t=1}^n \left[ h(Y_{j,t} | Y_j^{t-1}, S^n) - h(Y_{j,t} | W_0, W_1, \dots, W_\wp, Y_j^{t-1}, S^n) \right] + n\epsilon_n^{(j)}$$

$$\overset{(a)}{=} \sum_{t=1}^{n}\left[h(Y_{j,t}|Y_j^{t-1},S^n) - h(Y_{j,t}|W_0,W_1,\ldots,W_p,Y_j^{t-1},S^n,X_{1,t},\ldots,X_{p,t})\right] + n\epsilon_n^{(j)}$$

$$\overset{(b)}{\leq} \sum_{t=1}^{n}\left[h(Y_{j,t}|S_t) - h(Y_{j,t}|X_{1,t},\ldots,X_{p,t},S_t)\right] + n\epsilon_n^{(j)}$$

$$= \sum_{t=1}^{n} I(X_{1,t},\ldots,X_{p,t}; Y_{j,t}|S_t) + n\epsilon_n^{(j)}$$

(63)

where $\epsilon_n^{(j)} \to 0$ as $n \to \infty$. The equality in (a) is due to the fact that $(X_{1,t},\ldots,X_{p,t})$ is a deterministic function of $(W_0,W_1,\ldots,W_p,E_1,\ldots,E_p)$ and the inequality in (b) holds because conditioning does not increase entropy and also $(W_0,W_1,\ldots,W_p,Y_j^{t-1},S^{t-1},S_t^n) \to (S_t,X_{1,t},\ldots,X_{p,t}) \to Y_{j,t}$ forms a Markov chain. Note that for $i = 1,\ldots,p$, we have:

$$Pr\left(x_{i,t}|w_0,e_{i,t},\{x_{k,t}\}_{k=1,k\neq i}^{p}\right) = \sum_{w_i} Pr\left(x_{i,t},w_i|w_0,e_{i,t},\{x_{k,t}\}_{k=1,k\neq i}^{p}\right) = \sum_{w_i} Pr(w_i)Pr(x_{i,t}|w_0,e_{i,t},w_i).$$

(64)

Thus, the following Markov relationship holds:

$$\{X_{k,t}\}_{k=1,k\neq i}^{p} \to (\widetilde{U}_t,E_{i,t}) \to X_{i,t}, \qquad i = 1,\ldots,p$$

(65)

In this step, we introduce a time-sharing r.v. $Q$ uniformly distributed over the set $\{1,\ldots,n\}$ and independent of other r.v.'s. Define $X_i \triangleq X_{i,Q}$ and $E_i \triangleq E_{i,Q}$ for $i = 1,\ldots,p$, $Y_j \triangleq Y_{j,Q}$ and $Z_j \triangleq Z_{j,Q}$ for $j = 1,\ldots,p$, $S_{ji} \triangleq S_{ji,Q}$ for $i = 1,\ldots,p, j = 1,\ldots,q$, $\widetilde{U} \triangleq \widetilde{U}_Q$, and also the state matrix $S \triangleq S_Q$. We Note that:

$$Y_j = \sum_{i=1}^{p} S_{ji}X_i + Z_j, \qquad j = 1,\ldots,q$$

(66)

where $Z_j, j = 1,\ldots,q$, is a zero mean Gaussian r.v. with variance $\sigma_j^2$, which is independent of $Q, X_i, i = 1,\ldots,p$, and also the state matrix $S$. Therefore, from (62) and (63) we can write:

$$\begin{cases} \sum_{k\in\Lambda} R_k \leq I(\{X_k\}_{k\in\Lambda}; Y_j|\widetilde{U}_Q,Q,\{X_k\}_{k\notin\Lambda},S) + \epsilon_n^{(j,\Lambda)} \\ \sum_{k=0}^{p} R_k \leq I(X_1,\ldots,X_p; Y_j|S,Q) + \epsilon_n^{(j)} \leq I(Q,X_1,\ldots,X_p; Y_j|S) + \epsilon_n^{(j)} = I(X_1,\ldots,X_p; Y_j|S) + \epsilon_n^{(j)} \end{cases}$$

(67)

where the last equality is due to that $Q \to S, X_1,\ldots,X_p \to Y_j$ forms a Markov chain, as arises from (66). On the other hand, from the Markov relationships in (65), the following Markov chains result:

$$\{X_k\}_{k=1,k\neq i}^{p} \to (Q,\widetilde{U},E_i) \to X_i, \qquad i = 1,\ldots,p$$

(68)

Now by redefining $U \triangleq (Q,\widetilde{U})$ and by letting $n$ tends to infinity in (67), we obtain:

$$\begin{cases} \sum_{k \in \Lambda} R_k \leq I(\{X_k\}_{k \in \Lambda}; Y_j | U, \{X_k\}_{k \notin \Lambda}, S) \\ \sum_{k=0}^{p} R_k \leq I(X_1, \ldots, X_p; Y_j | S) \end{cases}$$

(69)

where the joint p.d.f. of the r.v.'s $\{X_i\}_{i=1}^{p}, \{E_i\}_{i=1}^{p}, U$ satisfies (29). Furthermore, by definition of the code, the transmitted codewords should satisfy the power constraints in (6). Thus, the bounds in (69) should be considered under the following power constraints:

$$\mathbb{E}[X_i^2] \leq P_i, \qquad i = 1, \ldots, p$$

(70)

This completes the proof. ∎

# APPENDIX III

## PROOF OF LEMMA 4

Consider a length-$n$ code with average error probability $P_e^{(n)}$, for the two-user Gaussian fading MAC with conferencing encoders, such that $P_e^{(n)} \to 0$ as $n \to \infty$. Define new r.v.'s $\tilde{U}_t, t = 1, \ldots, n$, as: $\tilde{U}_t \triangleq (V_1^K, V_2^K)$.

Using Fano's inequality we can write:

$nR_1 \leq H(W_1) = I(W_1; Y^n, S^n) + H(W_1 | Y^n, S^n)$

$\leq I(W_1; Y^n, S^n) + n\epsilon_n^{(1)} \leq I(W_1; Y^n, S^n, W_2) + n\epsilon_n^{(1)}$

$= I(W_1; Y^n, S^n | W_2) + n\epsilon_n^{(1)}$

$\leq I(W_1; Y^n, S^n, V_1^K, V_2^K | W_2) + n\epsilon_n^{(1)}$

$= I(W_1; Y^n, S^n | V_1^K, V_2^K, W_2) + I(W_1; V_1^K, V_2^K | W_2) + n\epsilon_n^{(1)}$

(71)

where $\epsilon_n^{(1)} \to 0$ as $n \to 0$. Now, for the first term in (71) we have:

$I(W_1; Y^n, S^n | V_1^K, V_2^K, W_2) = I(W_1; Y^n | V_1^K, V_2^K, W_2, S^n)$

$= \sum_{t=1}^{n} I(W_1; Y_t | V_1^K, V_2^K, W_2, S^n, Y^{t-1})$

$= \sum_{t=1}^{n} h(Y_t | V_1^K, V_2^K, W_2, S^n, Y^{t-1}) - h(Y_t | V_1^K, V_2^K, W_2, S^n, Y^{t-1}, W_1)$

$\stackrel{(a)}{=} \sum_{t=1}^{n} h(Y_t | V_1^K, V_2^K, W_2, S^n, Y^{t-1}, X_{2,t}) - h(Y_t | V_1^K, V_2^K, W_2, S^n, Y^{t-1}, W_1, X_{1,t}, X_{2,t})$

$\stackrel{(b)}{\leq} \sum_{t=1}^{n} h(Y_t | \tilde{U}_t, X_{2,t}, S_t) - h(Y_t | \tilde{U}_t, X_{1,t}, X_{2,t}, S_t)$

$$= \sum_{t=1}^{n} I(X_{1,t}; Y_t | \tilde{U}_t, X_{2,t}, S_t) \tag{72}$$

where the equality (a) is due to that $X_{2,t}$ is a deterministic function of $(W_2, V_2^K, E_{2,t} = \xi_2(S_t))$, and the inequality (b) is because conditioning does not increase entropy and also $Y_t$ conditioned on $(X_{1,t}, X_{2,t}, S_t)$ is independent of $(V_1^K, V_2^K, W_2, S^{t-1}, S_{t+1}^n, Y^{t-1}, W_1)$. Furthermore, by exactly the same procedure as in [7, p. 443], we have:

$$I(W_1; V_1^K, V_2^K | W_2) \leq nC_{12} \tag{73}$$

Combining (71)-(73), we obtain:

$$nR_1 \leq \sum_{t=1}^{n} I(X_{1,t}; Y_t | \tilde{U}_t, X_{2,t}, S_t) + nC_{12} + n\epsilon_n^{(1)} \tag{74}$$

Similarly one can show that:

$$nR_2 \leq \sum_{t=1}^{n} I(X_{2,t}; Y_t | \tilde{U}_t, X_{1,t}, S_t) + nC_{21} + n\epsilon_n^{(2)} \tag{75}$$

where $\epsilon_n^{(2)} \to 0$ as $n \to 0$. For the sum rate, we can write:

$$n(R_1 + R_2) \leq H(W_1, W_2) = I(W_1, W_2; Y^n, S^n) + H(W_1, W_2 | Y^n, S^n)$$

$$\leq I(W_1, W_2; Y^n, S^n) + n\epsilon_n^{(3)}$$

$$\leq I(W_1, W_2; Y^n, S^n, V_1^K, V_2^K) + n\epsilon_n^{(3)}$$

$$= I(W_1, W_2; Y^n, S^n | V_1^K, V_2^K) + I(W_1, W_2; V_1^K, V_2^K) + n\epsilon_n^{(3)} \tag{76}$$

where $\epsilon_n^{(3)} \to 0$ as $n \to 0$. Now, observe that:

$$I(W_1, W_2; Y^n, S^n | V_1^K, V_2^K) = I(W_1, W_2; Y^n | V_1^K, V_2^K, S^n)$$

$$= \sum_{t=1}^{n} I(W_1, W_2; Y_t | V_1^K, V_2^K, S^n, Y^{t-1})$$

$$= \sum_{t=1}^{n} h(Y_t | V_1^K, V_2^K, S^n, Y^{t-1}) - h(Y_t | V_1^K, V_2^K, S^n, Y^{t-1}, W_1, W_2)$$

$$= \sum_{t=1}^{n} h(Y_t | V_1^K, V_2^K, S^n, Y^{t-1}) - h(Y_t | V_1^K, V_2^K, S^n, Y^{t-1}, W_1, W_2, X_{1,t}, X_{2,t})$$

$$\leq \sum_{t=1}^{n} h(Y_t | \tilde{U}_t, S_t) - h(Y_t | \tilde{U}_t, X_{1,t}, X_{2,t}, S_t)$$

$$= \sum_{t=1}^{n} I(X_{1,t}, X_{2,t}; Y_t | \widetilde{U}_t, S_t) \tag{77}$$

and again similar to [7, p. 443]

$$I(W_1, W_2; V_1^K, V_2^K) \leq n(C_{12} + C_{21}) \tag{78}$$

Therefore, by combining (76)-(78) we obtain:

$$n(R_1 + R_2) \leq \sum_{t=1}^{n} I(X_{1,t}, X_{2,t}; Y_t | \widetilde{U}_t, S_t) + n(C_{12} + C_{21}) + n\epsilon_n^{(3)} \tag{79}$$

To derive the last bound on the sum rate in (53) we proceed as:

$$n(R_1 + R_2) \leq I(W_1, W_2; Y^n, S^n) + n\epsilon_n^{(3)}$$

$$= \sum_{t=1}^{n} I(W_1, W_2; Y_t | S^n, Y^{t-1}) + n\epsilon_n^{(3)}$$

$$= \sum_{t=1}^{n} h(Y_t | S^n, Y^{t-1}) - h(Y_t | S^n, Y^{t-1}, W_1, W_2) + n\epsilon_n^{(3)}$$

$$= \sum_{t=1}^{n} h(Y_t | S^n, Y^{t-1}) - h(Y_t | S^n, Y^{t-1}, W_1, W_2, X_{1,t}, X_{2,t}) + n\epsilon_n^{(3)}$$

$$\leq \sum_{t=1}^{n} h(Y_t | S_t) - h(Y_t | X_{1,t}, X_{2,t}, S_t) + n\epsilon_n^{(3)}$$

$$= \sum_{t=1}^{n} I(X_{1,t}, X_{2,t}; Y_t | S_t) + n\epsilon_n^{(3)} \tag{80}$$

By combining (74), (75), (79) and (80), applying a time-sharing argument similar to Appendix I, and then letting $n$ tends to infinity, the bounds in (53) are derived, where the fact that the underlying signaling satisfies the power constraints in (11) should also be considered. Finally, considering the code construction defined in Section II, one can verify that the following Markov relationships hold:

$$X_{2,t} \to (\widetilde{U}_t, E_{1,t}) \to X_{1,t},$$
$$X_{1,t} \to (\widetilde{U}_t, E_{2,t}) \to X_{2,t}$$

$$\tag{81}$$

The above Markov chains justify the distribution over which the union in (53) is taken. This completes the proof. ■